\documentclass[iop]{emulateapj}
\usepackage{graphicx,epsfig,amssymb,color,lineno}
\usepackage{natbib}
\usepackage{amsmath}
\usepackage{enumitem}
\usepackage{cleveref}
\usepackage{gensymb}

\slugcomment{\footnotesize }
\shorttitle{Lyman Continuum Escape Fraction at $z\sim1.3$}
\shortauthors{Alavi et al.}

\newcommand{\RN}[1]{%
  \textup{\uppercase\expandafter{\romannumeral#1}}%
}

\begin{document}

\title{Lyman Continuum Escape Fraction from Low-mass Starbursts at \lowercase{z}$=1.3$\footnotemark[*]}\footnotetext[*]{Based on observations made with the NASA/ESA Hubble Space Telescope, 
obtained from the data archive at the Space Telescope Science Institute. STScI is operated by the Association of Universities for Research in Astronomy, Inc. under NASA contract NAS 5-26555.}

\author{\sc Anahita Alavi\altaffilmark{1}, James Colbert\altaffilmark{1}, Harry I. Teplitz\altaffilmark{1}, Brian Siana\altaffilmark{2}, 
Claudia Scarlata\altaffilmark{3}, Michael Rutkowski\altaffilmark{4}, Vihang Mehta\altaffilmark{3}, Alaina Henry\altaffilmark{5}, Y. Sophia Dai \altaffilmark{6}, Francesco Haardt\altaffilmark{7}, Micaela Bagley\altaffilmark{8}}

\altaffiltext{1}{IPAC, California Institute of Technology, 1200 E. California Boulevard, Pasadena, CA 91125, USA, anahita@ipac.caltech.edu}
\altaffiltext{2}{Department of Physics and Astronomy, University of California, Riverside, CA 92521, USA}
\altaffiltext{3}{Minnesota Institute for Astrophysics, University of Minnesota, Minneapolis, MN 55455, USA}
\altaffiltext{4}{Department of Physics and Astronomy, Minnesota State University, Mankato, MN 56001, USA}
\altaffiltext{5}{Space Telescope Science Institute, 3700 San Martin Drive, Baltimore, MD, 21218, USA}
\altaffiltext{6}{Chinese Academy of Sciences South America Center for Astronomy (CASSACA)/National Astronoical Observatories of China (NAOC), 20A Datun Road, Beijing 100012, China}
\altaffiltext{7}{DiSAT, Universi\'a degli Studi dell'Insubria, 22100 Como, Italy} 
\altaffiltext{8}{Department of Astronomy, The University of Texas at Austin, Austin, TX 78712, USA}

\begin{abstract}
We present a new constraint on the Lyman Continuum (LyC) escape fraction at $z\sim1.3$. 
We obtain deep, high sensitivity far-UV 
imaging with the Advanced Camera for Surveys (ACS) Solar Blind Channel (SBC) on the Hubble Space Telescope (HST), targeting 11 
star-forming galaxies at $1.2<z<1.4$.
The galaxies are selected from the 3D-HST survey to have high H$\alpha$ equivalent width (EW) with EW $> 190$ \AA, 
low stellar mass (M$_{*} < 10^{10} \ \text{M}_{\odot}$) and U-band magnitude of U$<24.2$. 
These criteria identify young, low metallicity star bursting populations 
similar to the primordial star-forming galaxies believed to have reionized the universe. 
We do not detect any LyC signal (with S/N $ >3$) in the individual galaxies or in the stack in the far-UV images. 
We place $3\sigma$ limits on the relative escape fraction of individual 
galaxies to be $f_{esc,rel}<[0.10-0.22]$ and a stacked $3\sigma$ limit of $f_{esc,rel}<0.07$. 
Comparing to the confirmed LyC emitters from the literature, the galaxies in our sample span similar ranges 
of various galaxy properties including stellar mass, dust attenuation, and star formation rate (SFR). 
In particular, we compare the distribution of H$\alpha$ and [OIII] EWs of confirmed LyC emitters and non-detections including the galaxies in this study. 
Finally, we discuss if a dichotomy seen in the distribution of H$\alpha$ EWs can perhaps distinguish the LyC emitters from the non-detections.
\end{abstract}


\section{Introduction}
Reionization is the last major phase transition in the universe, when Lyman Continuum (LyC) photons ionized the Inter-Galactic Medium (IGM). 
Hundreds of hours of observations with the leading facilities in the world have been devoted to understanding when and how reionization happened. 
From observations studying the duration of reionization, it is now believed that the reionization era 
ended by redshift $\sim 6.$ This has been demonstrated by various methods including the 
Gunn--Peterson effect in the spectra of high-redshift QSOs \citep[e.g., ][]{fan06} and the downturn seen in the fraction of 
Ly$\alpha$ emitters among Lyman break galaxies beyond z=6 \citep[e.g., ][]{stark10,pen14,mas19}. In addition, the \citet{planck16} estimate an average redshift between 7.8 and 8.8 for this epoch, 
adopting an instantaneous reionization model.

 However, despite a considerable number of studies, it remains uncertain which sources dominated the emission of LyC photons during the reionization epoch. 
 Both active galactic nuclei (AGN) and young massive stars produce LyC photons and can contribute to the reionization of the IGM. 
 Though, there are some studies indicating that AGNs might be important  \citep{gial15,gial19}, many studies have concluded that they can not be the 
 primary contributors to reionization \citep[e.g., ][]{willot05,sia08a,masters12,mat18,par18,kul19}.

In the absence of AGNs, the young massive stars in star-forming galaxies seem to be the primary 
 sources of LyC photons in the early universe. Recent studies 
 \citep[e.g., ][]{bou17,liv17,ate18,ish18,ono18,pel18,yue18} of UV luminosity functions at $z>6$ have found steep faint-end slopes suggesting a 
 large number density of faint star-forming galaxies at these redshifts. These findings suggest that faint star-forming galaxies may 
 play a significant role during the reionization era \citep[e.g., ][]{rob13,bou15b,mas15, rob15}. 
In particuler, (1) faint galaxies must produce sufficient LyC photons and 
(2) these photons must escape absorption within interstellar medium (ISM) and reach the IGM. 
The former point can be expected as the faint star-forming galaxies are the most abundant galaxies particularly in the early universe. 
However, the ionizing photon production rate of these faint galaxies still requires investigation \citep{emami20}.
To quantify the latter point, we need to measure 
the fraction of LyC photons that escape (i.e., $f_{\text{esc}}$) the ISM and reach IGM. 

A direct measure of $f_{\text{esc}}$ at $z>4$ is difficult due to the high opacity of IGM \citep{ino14}. Therefore, direct study of 
escaping LyC photons is limited to low and intermediate redshifts. Early studies of the LyC emission escaping nearby galaxies \citep{lei95,deh01,gri09}
resulted only in upper limits, suggesting $f_{\text{esc}}$ of a few percent.  
At higher redshifts ($z\sim 3$), early studies \citep{sha06,nes11} reported detections, 
but several of those detections turned out to be contaminated by
foreground sources at lower redshifts \citep{van10b,van12,sia15}. Later on, more studies found low redshift contaminants 
in their sample of LyC emitter candidates \citep{van12, mos15, gra16}.
Additional high-z studies found null detections and only obtained upper limits 
\citep{sia07,iwa09,sia10,bouts11,rut16,rut17,smi18}. 

Recently, several LyC galaxies have been detected at low \citep[$z<0.4$;][]{lei13,bor14,leith16,izo16a,izo16b,izo18a,izo18b} 
and high redshifts \citep[z= 2-4;][]{sha16,van16,nai17,bia17,ste18,van18,fle19,riv19}. 
However, all attempts at intermediate redshifts of $z\sim1$ have given null results \citep{mal03,sia07,cow09,bri10,sia10,rut16}. 
Although the rate of success in finding LyC emitters at all redshifts has been low, some detections at $z\sim1$ should have been expected. 
In fact, there are two reasons that one might expect the detection of LyC emission at $z=1-2$ to be easier than at low redshifts:
(1) $z\sim1$ star-forming galaxies at fixed stellar-masses 
have higher star formation rates than $z\sim0$ galaxies \citep[and references therin]{mad14} and (2) high-redshift star-forming galaxies 
have less dust than their low-redshift analogs \citep[e.g., ][]{red06}. 
The former favors a higher production rate of LyC photons in higher redshift galaxies \citep[see also][]{matt17}. 
The latter results in less absorption of LyC photons as they travel through the ISM to IGM.


The recent detections of LyC emission have been accomplished with the combination of high sensitivity 
observations and apparently effective selection techniques. 
One such technique identifies galaxies with high ratios of [OIII]/[OII], which is potentially associated with density-bounded HII regions. 
 \citet{nak14} present a photoionization model calculation with CLOUDY \citep{ferland98} that suggests that galaxies with 
 high [OIII]/[OII] ratios are good candidate high $f_{\text{esc}}$ objects. They also show that their finding 
 is consistent with the ratio of [OIII]/[OII] $\sim 1-4$ of two LyC leakers \citep{lei11,lei13} known at that time.
 The potential of this criterion to identify LyC leakers was later
 investigated in several low-redshift studies of galaxies with [OIII]/[OII] $\gtrsim 5$ \citep{izo16a,izo16b,izo18a,izo18b}, which were successful in finding LyC emitters using the 
 Cosmic Origins Spectrograph (COS) on HST. 
 At high redshift, the first LyC emitter that was discovered, $Ion2$, was 
 found to have a high ratio of [OIII]/[OII] $\gtrsim10$ \citep{van16}. In addition, \citet{fai16} find a 
 positive correlation between the [OIII]/[OII] ratio and $f_{\text{esc}}$ compiling thirteen detections and upper limits from the literature. 
 In contrast, \citet{stasinska15} argue that this line
 ratio on its own is not a sufficient diagnostic tool for LyC leakage. 
 This was later validated by several unsuccessful searches of LyC leakage among galaxies with high [OIII]/[OII] ratio \citep{rut17,nai18} and a statistical analysis by \citet{izo20}.
 
 Relatedly, many of the confirmed LyC emitters at high \citep{van16,nai17,fle19} and low redshifts 
 \citep{izo16a,izo16b,izo18a,izo18b} also display extreme [OIII] EWs. 
 Indeed, extreme [OIII] emitters at $z=0.1-0.3$, known as ``Green Pea" galaxies \citep{cardamone09}, have long 
 been studied as potential candidates for high LyC escape fraction 
 \citep{jas13, nak14,hen15,izo16a,izo16b,izo18a,izo18b}. However, recently, \citet{nai18} found no 
 LyC detection for their sample with high [OIII] EWs, thus
  raising a question about the reliability of extreme [OIII] EW as an effective tracer of LyC emission. 
  Similar conclusions are also found for individual 
 non-detections at low redshifts by \citet{izo17} and high redshifts by \citet{amo14} and \citet{vas16}.


In this paper, we search for LyC photons in low-mass emission line galaxies during a 
burst in their star formation. We select galaxies to have strong H$\alpha$ emission lines with rest-frame EW$ > 190$ \AA. The H$\alpha$ line is an indicator of instantaneous SFR, and thus it traces 
young and hot O-type stars, which are responsible for the LyC production in galaxies. To this end, we conduct a deep far-UV 
imaging program and exploit the high sensitivity and high spatial resolution of the SBC of the ACS \citep{for98} onboard HST. Our observations would be sensitive to LyC photons 
at $z\sim1.3$, a redshift from which there has been no detections of escaping ionizing radiation to date.


This paper is organized as follows. 
Section \ref{sec:observation_and_data} presents the HST observations, reduction of the data and sample selection.
 In Section \ref{sec:photometry}, we discuss the steps involved in measuring the observed far-UV photometry.
In Section \ref{sec:results}, we show the results including upper limits to the LyC fluxes of individual galaxies and stacks, and we calculate the 
upper limit of the escape fraction of ionizing photons. Section \ref{sec:discussion} compares our study and other LyC efforts in the literature to better understand the 
galaxy properties that favor LyC leakage.

Throughout the text, we use a flat $\Lambda$CDM cosmology with $H_{ 0}=70$ km s$^{-1}$ Mpc$^{-1}$, $\Omega_{M}=0.3$, and $\Omega_{\Lambda}=0.7$. All 
magnitudes are in the AB system \citep{oke83}. 
  

\section{Observation and Data Reduction}
\label{sec:observation_and_data}

\subsection{Sample Selection}
We targeted 11 star-forming galaxies detected by the 3D-HST survey \citep[PI: van Dokkum;][]{bram12,mom15} of the 
GOODS-South and COSMOS fields with spectroscopic redshifts between $1.2<z<1.4$. 
The lower end of the redshift range is selected to avoid contamination from non-ionizing UV photons redward of the Lyman limit at 
912 \AA\ and the upper end ensures high sensitivity to Lyman Continuum photons (see Figure \ref{fig:filter}).

\begin{figure}
\epsscale{2.0}
\includegraphics[trim=0.2cm 0.1cm 0.1cm 0cm,clip=true, angle=0, width=\columnwidth]{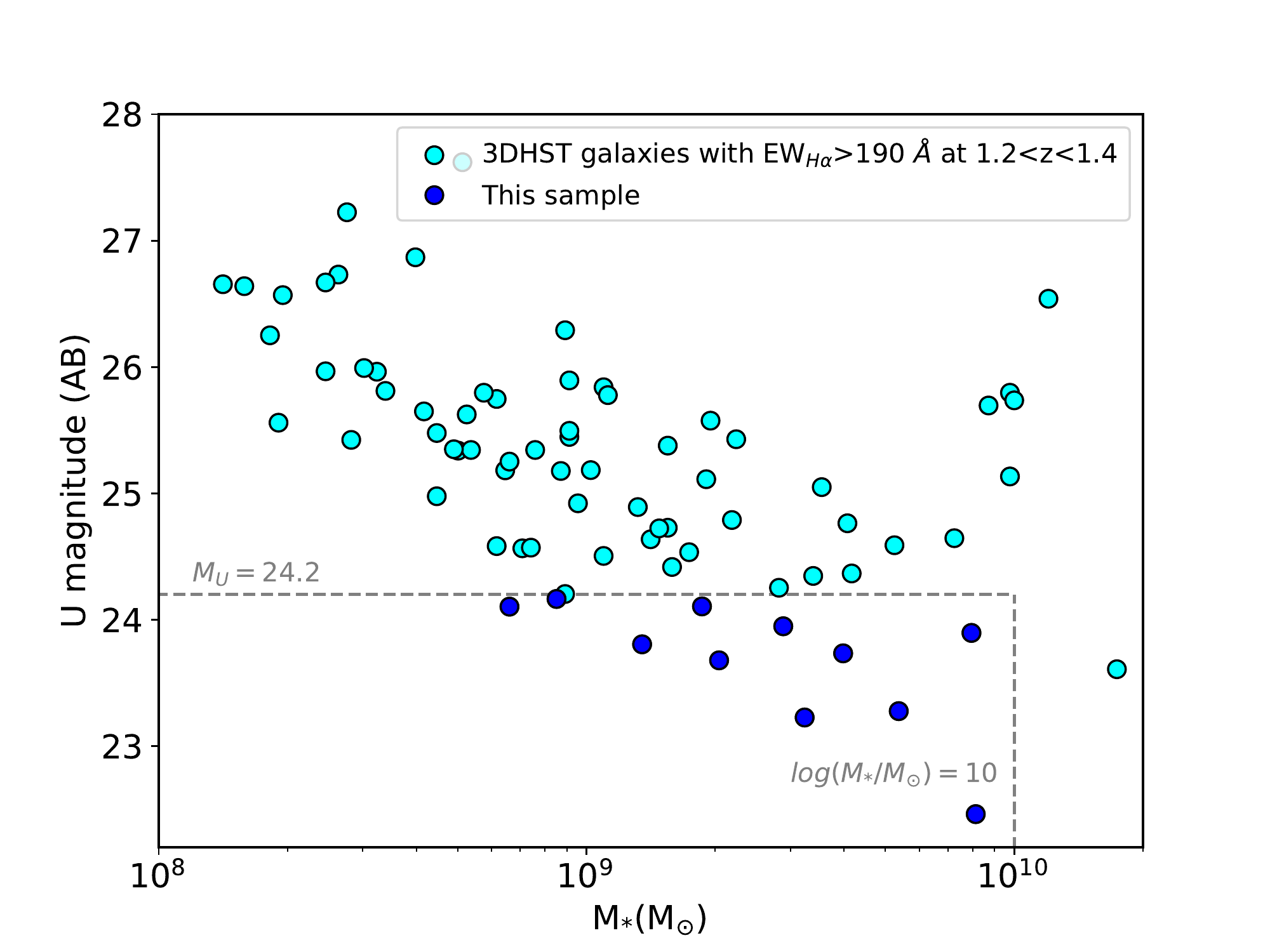}
\caption{The U-band magnitude versus stellar mass for all of the galaxies with $1.2<z<1.4$ and EW(H$\alpha)>190$ \AA\ in 
the 3D-HST catalogs. The dashed lines show our 
U-band magnitude and stellar mass cuts at $U=24.2$ and $\log (M_{*}/M_{\odot}) < 10$, respectively.
Our final sample consist of 11 sources, which happened to be located in the COSMOS and GOODS-South fields specified as blue circles.}
\label{fig:selection}
\end{figure}

To select targets from the 3D-HST spectra, we used custom measurements \citep{rut16}, made with code originally developed 
for the {WFC3 IR Spectroscopic Parallel} survey \citep[WISP; PI:Malkan;][]{ate10}. 
Galaxies are selected to have strong H$\alpha$ emission lines with rest-frame EW $>190$ \AA\ and stellar masses of 
$\log (\text{M}_{*}/\text{M}_{\odot}) < 10$. Therefore, our sample includes young, low metallicity, 
and low-mass star-bursting populations, similar to the class of galaxies believed to reionize the universe. We note that the lower mass selection 
criterion only excludes two galaxies from our sample as the extreme emision-line selected galaxies (i.e., high EW) have preferentially 
 low stellar continuum and hence low stellar masses. 
 We further require that these galaxies are bright enough that they can provide a meaningful limit to LyC escape fraction, with 
 U-band$<24.2$ in the CFHT MegaCam U-band \citep{erben09,hil09} or the VLT VIMOS U-band \citep{noni09} images. 
 As a result of this criterion, the selected galaxies have UV luminosities similar to $L^{*}_{UV}$ at z$=1.3$ 
 \citep[using $L^{*}_{UV}$ estimate from][]{ala16}.
 Figure \ref{fig:selection} summarizes our 
selection criteria.  We plot all of the high H$\alpha$ 
 EW galaxies within 3D-HST, and show the UV magnitude and stellar mass cut that selects appropriate targets.
A list of targets is given in Table \ref{tab:objects}.

\subsection{Observations}
\label{sec:observation}
\begin{figure}
\epsscale{2.0}
\includegraphics[trim=0.2cm 0.1cm 0.1cm 0cm,clip=true, width=\columnwidth]{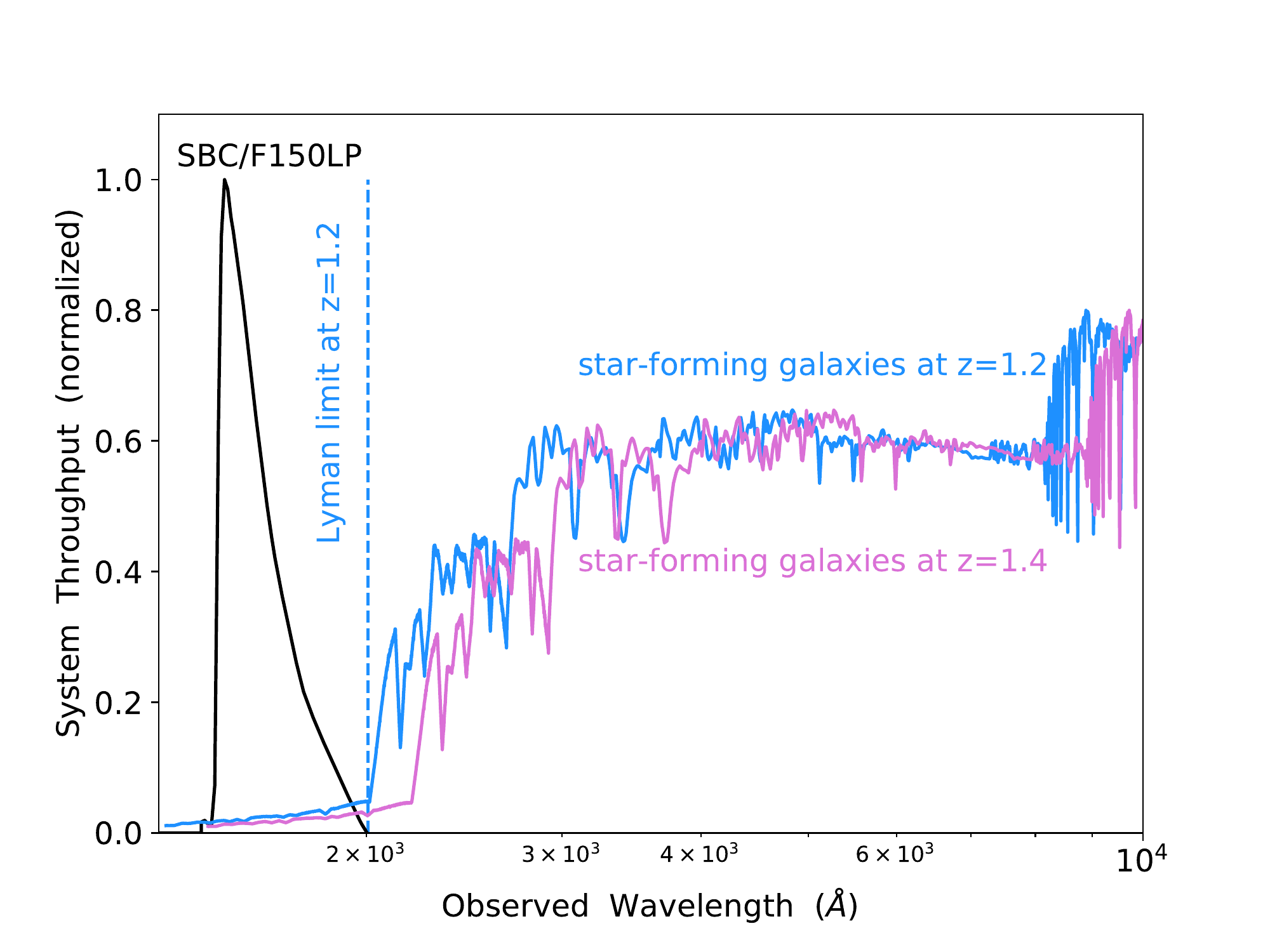}
\caption{Total system throughput for the ACS SBC/F150LP filter. Two model SEDs of star-forming galaxies at z=1.2 (solid blue) and 1.4 (solid purple), 
the low and high cuts of our redshift range, are also shown to 
demonstrate that no non-ionizing flux is entering this system throughput. The SEDs are from BC03 synthetic stellar 
population models  \citep{bru03}  representing of normal star-forming galaxies at $z\sim1$ with 
constant SFR, solar metallicity, 100 Myr age and E(B-V)=0.1. The dashed blue line shows the Lyman limit cut at rest-frame 912 \AA\ at $z=1.2$. }
\label{fig:filter}
\end{figure}

The goal of this work is to search for escaping LyC photons from strong emission-line galaxies at $1.2<z<1.4$. 
To this end, we obtained far-UV imaging (program ID 14123, PI: J. Colbert) of our targets using the F150LP filter of 
SBC on ACS \citep{for98}. This filter has significant transmission in the wavelength range of 
$1450<\lambda < 2000$ \AA\ (see Figure \ref{fig:filter}). The blue cutoff avoids the contamination from the geocoronal emission lines
(Ly$\alpha$ and OI lines at 1304 and 1356 \AA), which would significantly increase the background in the images. The red cutoff 
is dictated by the decreasing sensitivity of the Multi-Anode Microchannel Array (MAMA) toward redder wavelengths \citep{sia10}.
The effective wavelength of this filter 
is $\lambda_{eff}=1616.67$ \AA, which probes LyC photons at $\lambda_{rest}\sim700$ \AA\ at $z=1.3$. 
We note that most LyC searches including spectroscopic \citep[COS observations of local LyC emitters in][]{izo16a,izo16b,izo18a,izo18b} and photometric studies \citep[WFC3/UVIS observations of high-redshift LyC emitters in][]{van16,nai17,fle19}, are sensitive to LyC flux at 900 \AA. The neutral hydrogen opacity at 
 $\lambda_{rest}\sim900$ \AA\ is about twice of that at $\lambda_{rest}\sim700$ \AA, because the photoionization cross section decreases as $\nu^{-3}$.

The MAMA detector of SBC is a photon-counting device and is not affected by cosmic rays. 
The detector has no read noise and its primary source of noise is dark current. The SBC dark current has two components. 
The first is a steady, spatially uniform count rate that does not change with 
the detector temperature. It is estimated to be a very stable value of $8.11\times10^{-6}$ cts/pix/s \citep{avi17}. 
The second component is a temperature-dependent glow that rises at $T>25 \ {\celsius}$. 
In previous studies \citep{tep06,sia07,avi17}, it has been noted that the variable dark glow is most prominent near the center 
of the chip and the lower left corner of the detector has a stable dark rate even at elevated temperatures. 
Therefore, we designed our observations to place all of our targets in the detector corner least affected by 
the central glow (position $x=250$, $y=250$ pixel on the SBC detector), similar to the strategy of \citet{sia10}.

Each target is imaged in one visit to a varying depth of one, two or three orbits depending on the galaxy's rest-UV brightness 
(also see Section \ref{subsec:escape fraction}). Each orbit consists of four dithered exposures with similar exposure times of 654s or 692s.

 \begin{deluxetable*}{ccccccccc}{!t}
\tablecaption{SBC Sample}
\tablewidth{0pt}
\tablehead{\colhead{ID} & \colhead{RA [deg]} & \colhead{Dec [deg]} & \colhead{z} &  \colhead{EW$_{rest} (H\alpha$)[\AA]} & \colhead{EW$_{rest}$([OIII])[\AA]} & \colhead{$\log$(M$_{*}$/M$_{\odot}$)\tablenotemark{a}}
 & \colhead{A$_{v}$\tablenotemark{a}} & \colhead{$\log$(SFR)[$M_{\odot} \ yr^{-1}$] \tablenotemark{a}}}\\
\startdata 
cosmos-1-111    &     150.07448        &       2.4172121       &       1.254       &      193.0       &        -*      &       9.31       &      0.4       &           -0.30  \\
cosmos-3-69 	&	150.09259	&	2.3244885	&	1.400       &       201.1       &      45.6         &        9.60       &        0.5       &        0.61 \\
cosmos-3-113   &	150.08089	&	2.3239765	&	1.257       &       190.7       &      160.1       &        9.27       &        0.5       &        0.45 \\
cosmos-7-64   &	150.06325	&	2.4488557	&	1.320       &       228.8       &      212.9       &        9.51       &        0.1       &        -0.79 \\
cosmos-13-80   &	150.12906	&	2.2307920	&	1.230       &       199.1       &      172.2       &        9.46       &        0.6       &        0.64 \\
cosmos-28-132   &	150.10593	&	2.4162436	&	1.262       &       498.6       &      419.7       &        8.95       &        0.1       &        -0.04 \\
goodss-6-124   &	53.199558	&	-27.863221	&	1.231       &       248.1       &      485.9       &        8.82       &        0.0       &        0.42 \\
goodss-9-108   &	53.062504	&	-27.764822	&	1.232       &       193.0       &      197.7          &        9.13       &        0.1       &        0.31 \\
goodss-21-24   &	53.198181	&	-27.878668	&	1.253       &       226.6       &      110.6       &        9.91       &        0.3       &        1.09 \\
goodss-27-124   &	53.193836	&	-27.844275	&	1.237       &       344.4       &      738.1      &        8.93       &        0.0       &        0.27 \\
goodss-30-67  	&	53.099816  	&	-27.730301	&	1.309       &       260.5       &      205.4       &        9.73       &        0.2       &        0.83 \\
\enddata 
\tablenotetext{a}{ All physical properties are from our SED fitting.}
\tablenotetext{*}{The [OIII] line for this source is very noisy and the line-fitting is uncertain. Therefore, we do not use the [OIII] EW estimate for this source.  }
\label{tab:objects}
\end{deluxetable*}

\subsection{Data reduction}
\label{sec:reduction} 
We downloaded the raw data from the Barbara A. Mikulski Archive for Space Telescopes (MAST), which we then processed 
using the PYRAF/STSDAS \texttt{CALACS} program to subtract the dark current and flat field the images. 
We use the dark image provided by the STScI as a reference to remove 
the primary dark component. This dark component does not account for the central glow which rises with temperature. This excess dark component 
is later subtracted as explained in Section \ref{sec:dark}. The \texttt{Astrodrizzle} code 
was then used to combine the exposures and make the final drizzled image. Here, we weight the individual frames by their exposure time 
and we set the output pixel scale to 0.03\arcsec and the pixfrac to 1. Because the SBC data have little sky background and are insensitive 
to cosmic rays, the sky subtraction and cosmic ray rejection steps have been turned off in running the \texttt{Astrodrizzle} code.

The images are drizzled and aligned to the CANDELS F606W tiles. 
The relative astrometry between SBC and reference images is always better than 1-1.5 pixels (i.e., the rms of the best alignment fit). 
We note that the popular \texttt{Tweakreg} code for the HST 
image alignment fails due to lack of an adequate number of compact, bright sources. 
Therefore, we manually identify matching sources 
on the individual science and reference images, and then use the \texttt{PYRAF} \texttt{geomap} code to calculate the shift between the images. 

We note that we could only align the images of 7 sources. For 
the remaining 4 (with IDs cosmos-7-64, cosmos-9-108, goodss-21-24, and goodss-30-67), no sources were detected in the 
individual SBC/F150LP images to be used for the alignment. Unless otherwise noted, we 
drop these 4 objects from the LyC analyses presented in this paper. 
Similar to the sample of 7 sources that we will be discussing, we did not detect any LyC flux for these 4 objects in the F150LP images.

\subsection{Dark subtraction}
\label{sec:dark}
As stated above, dark current is the dominant source of noise in these observations and a careful 
treatment of dark subtraction and its varying component is vital. Here, the total dark is a sum of the 
primary and excess dark components. We subtract these dark components separately in two stages.\
 
First, we subtract the primary calibration dark reference file \footnote{ We downloaded the reference file $04k1844aj\_drk$ 
 from the HST Calibration Reference Data System (CRDS).}, 
 which accounts for the low and stable dark current of $8.11 \times 10^{-6} \ \text{counts} \ \text{s}^{-1} \ \text{pix}^{-1}$ when the instrument is $<25 \ \celsius$. As explained in Section \ref{sec:reduction}, the primary dark subtraction is part of the image processing 
 done by the \texttt{CALACS} program.\
 
Second, although each target is located in the least affected region on the detector, we check for additional dark 
 current associated with the central glow. First from a visual inspection of different exposures in the image of each target, 
 we define a border separating the corner with stable background from the central region with varying dark glow. 
 We exclude those regions of the image where there are sources using a segmentation map from a \texttt{SExtractor} \citep{ber96} run on F606W images. 
 Using the F606W photometric aperture of each target (see Section \ref{sec:photometry}), we then 
 generate random apertures within the corner of each image and investigate the distribution of total flux within these regions. 
If the primary calibration dark image were adequate, we would expect each flux distribution to be centered at zero. 
However, these flux distributions are centered around non-zero values varying between $[5.0,11.0] \times 10^{-6} \ \text{counts} \ \text{s}^{-1} \ \text{pix}^{-1}$.
This is evidence of an excess dark current with a value of $\pm 30\%$ of the primary dark current (i.e., excess dark $=(1\pm 0.3)\times $primary dark). 
 Therefore, for each drizzled image, we subtract the median of the flux distribution of the random apertures within the corner. 
  We use these improved drizzled images for our LyC analyses.
 
This excess dark current appears to increase with the detector temperature.
 We also investigate the possibility of a gradient in the excess dark and we find that it rises from corner edge toward the center by about $30\%$. However, this dark gradient 
 does not affect our photometry because our sources are compact. Overall, these behaviors are consistent with those reported in \citet{tep06}, \citet{sia07}, and \citet{avi17}. However, 
 our estimate of the excess dark current in the corner is much weaker than what these studies have reported for the 
 central dark glow, thanks to our mitigation strategy.


 \section{photometry}
 \label{sec:photometry}
 For the UV, optical and Near-IR photometry, we use the public 3D-HST catalogs of COSMOS and 
 GOODS-South fields \citep{bram12,ske14}. They assembled the catalogs using a 
 combination of three Wide Field Camera 3 \citep[WFC3;][]{mac10} bands of F125W, F140W and F160W for detection and the PSF-matched HST 
 images of each field.
 For each galaxy, we use ground-based U-band ($\lambda_{rest} \sim 1600$ \AA) from CFHTMegaCam or VLT VIMOS and HST/SBC F150LP 
 images to compute the observed non-ionizing and ionizing UV (i.e., LyC) fluxes, respectively. 
 
 To define apertures for LyC measurements, 
 we started with the SExtractor segmentation maps of the 3D-HST catalogs. However, these photometric
 apertures are much larger than the area where we expect a significant 
 rest-frame far-UV flux from each galaxy. Therefore, we define new isophotal apertures using shorter wavelength filters, 
 which probe the rest-frame near-UV and thus areas of ongoing star formation in galaxies. Ideally, 
 we would use the U-band images as our detection band but they are low-resolution ground-based data and 
 require degrading the SBC images. We therefore choose to run the SExtractor in dual image mode using the 
 optical F606W image, which corresponds to rest-frame $\lambda \sim 2600$ \AA, for detection. We note that this is the 
 same Sextractor measurement that was referred to in Section \ref{sec:dark}. Because the F606W image is deep 
 (i.e, with $5\sigma$ depth of 28.3 and 29.4 magnitudes for COSMOS and GOODS-South, respectively), the isophotes are large. 
 Using a solution discussed in \citet{sia07}, we find that if we shrink the isophotes to include 80\% of the F606W total flux, the 
 area decreases by a factor of 1.5-2.7. This increases our far-UV sensitivity by 0.2-0.5 magnitude. 
 
 We calculate the $3\sigma$ upper limits of the LyC fluxes in two ways: \

\begin{itemize}

\item As explained above, the dark current is the dominating component of noise in the SBC images. 
We estimate the total dark current from the sum of the primary dark and excess dark component (see section \ref{sec:reduction}) within the isophotal aperture of 
each target. The total noise is then calculated  as $\sqrt{ \text{Total Dark}\times \text{Exposure time}\times \text{Area}_{\text{isophot}} }$. \\
\item We first use the F606W isophotal segmentation map to exclude all objects from the SBC images. We then randomly move the isophotal aperture of each target within the image corner where it is located and measure the flux 
within the aperture.
The final noise is then derived from the standard deviation of the distribution of random aperture fluxes.
\end{itemize}

Our estimates of the limits from these two measurements agree within $10\%$. Unless otherwise noted, 
we use the limits from the second technique as listed in Table \ref{tab:limits}.
 
 Finally, we correct the photometry for Galactic extinction using the values of $A_{V}=0.051 \ \& \ 0.021$ mag for 
 COSMOS and GOODS-South, respectively, as reported in \citet{ske14}. 
 These Galactic dust extinction values are based on the recalibration by \citet{sch11} of the COBE/DIRBE and IRAS/ISSA 
 dust maps. Assuming a Cardelli extinction curve \citep{car89} with $R_{V}=3.1$, 
 we estimate the $A_{1600}=0.13 \ \& \ 0.05$ mag, respectively.

 \section{Results}
 \label{sec:results}
 
 \subsection{Individual galaxies}
 \label{Individual galaxies}
\begin{figure*}
\centering
\includegraphics[angle=0,trim=0cm 0cm 0cm 0cm,clip=true,width=\columnwidth]{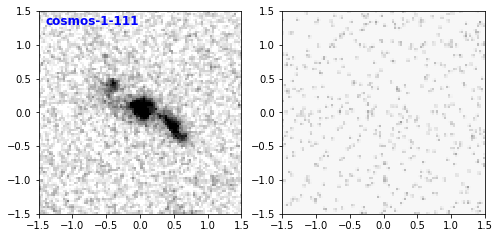}
\includegraphics[angle=0,trim=0cm 0cm 0cm 0cm,clip=true,width=\columnwidth]{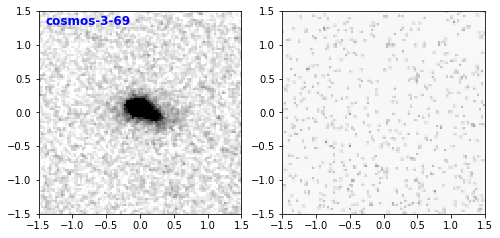}
\includegraphics[angle=0,trim=0cm 0cm 0cm 0cm,clip=true,width=\columnwidth]{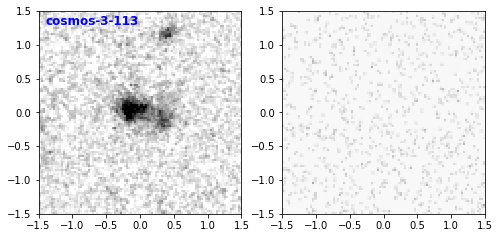}
\includegraphics[angle=0,trim=0cm 0cm 0cm 0cm,clip=true,width=\columnwidth]{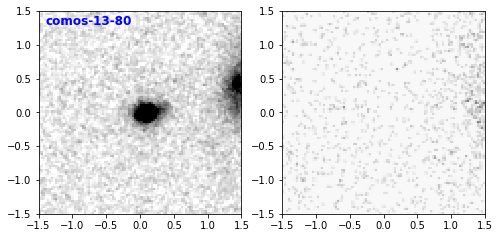}
\includegraphics[angle=0,trim=0cm 0cm 0cm 0cm,clip=true,width=\columnwidth]{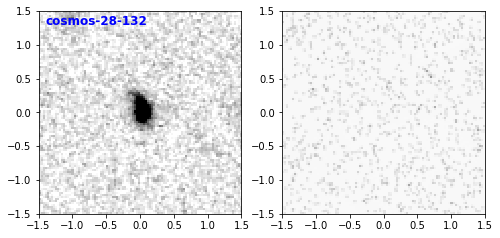}
\includegraphics[angle=0,trim=0cm 0cm 0cm 0cm,clip=true,width=\columnwidth]{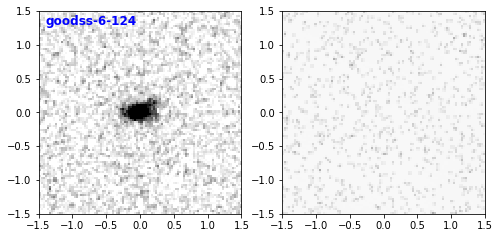}
\includegraphics[angle=0,trim=0cm 0cm 0cm 0cm,clip=true,width=\columnwidth]{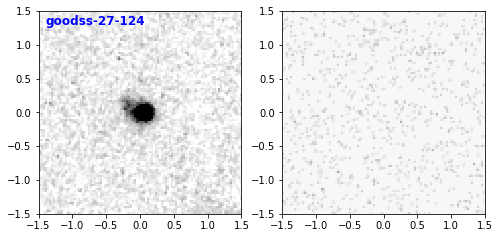}

\caption{The postage stamps of CANDELS F606W (left) and far-UV F150LP (right) of our targets. Each image is 3$\arcsec$ on a side. Most of these galaxies 
(except cosmos-1-111 and cosmos-3-113) display small and compact sizes. As seen in these images, none of these galaxies are detected in the F150LP images.} 
\label{fig:stamp}
\end{figure*}

\begin{figure}
\epsscale{2.0}
\includegraphics[trim=0.2cm 0.1cm 0.1cm 0cm,clip=true, width=\columnwidth]{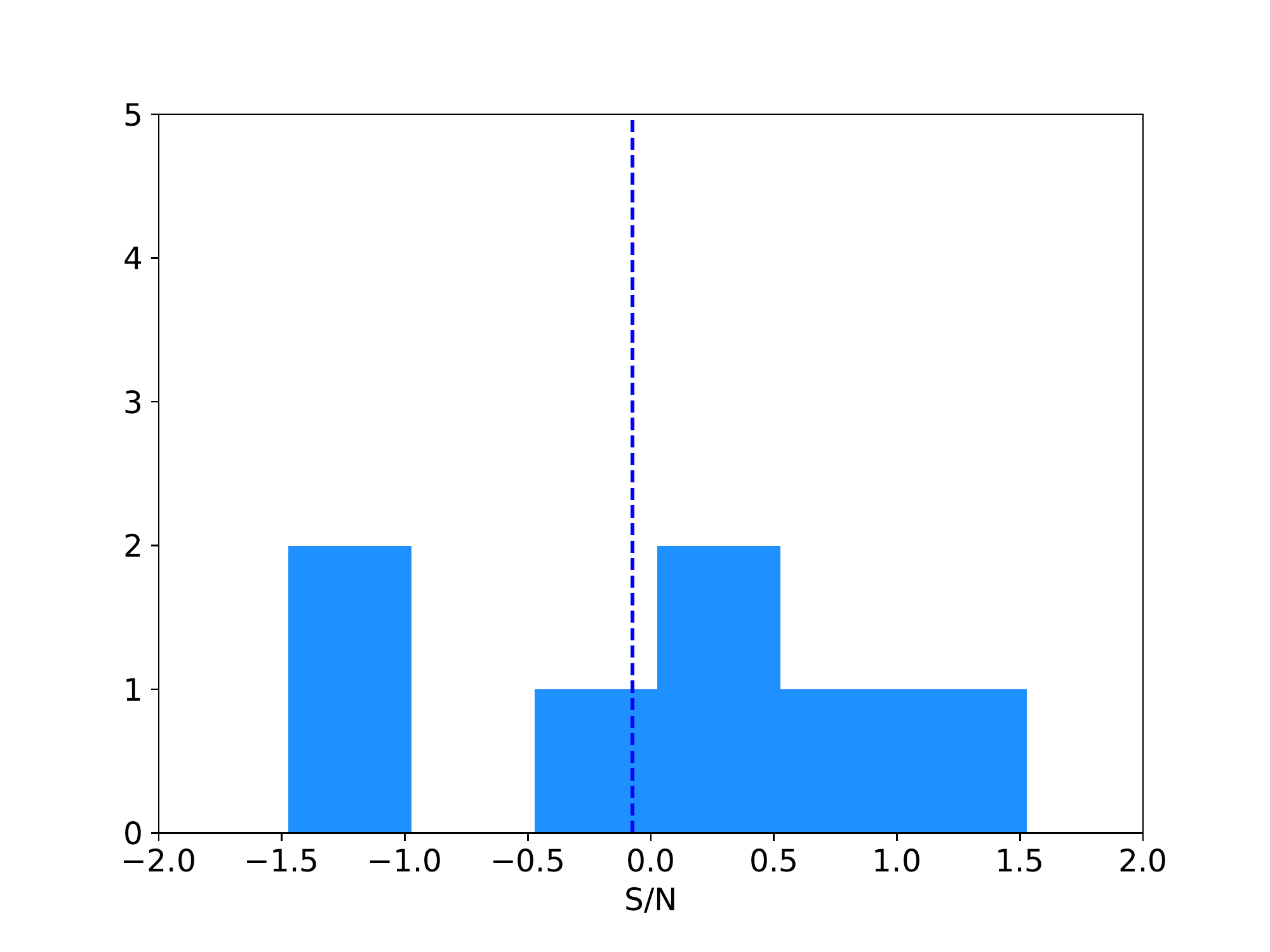}
\caption{The distribution of measured S/N of LyC fluxes for our sample of 7 galaxies. There is no detection above S/N$>3$. 
The distribution is nearly centered at zero with the dashed line indicating the arithmetic mean value at -0.07. }
\label{fig:sn}
\end{figure}

{\it We do not detect any individual galaxy with $S/N >3$ in the SBC LyC images.} Figure \ref{fig:stamp} shows the postage stamps of 
SBC far-UV and F606W images of each of the targets. The distribution of S/N values calculated using the second technique is shown in Figure \ref{fig:sn}. 
The distribution of the S/N is centered around zero with a mean value of -0.07 and standard deviation of 0.9.  We also 
search for LyC flux that may exist offset (up to 2\farcs0) from the UV continuum as discussed in 
\citet{iwa09} and \citet{nes13}.

 \subsection{Stack}
 \label{sec:stack}
\begin{figure}
\epsscale{1.0}
\includegraphics[trim=0.2cm 1cm 0.1cm 0cm,clip=true, width=\columnwidth]{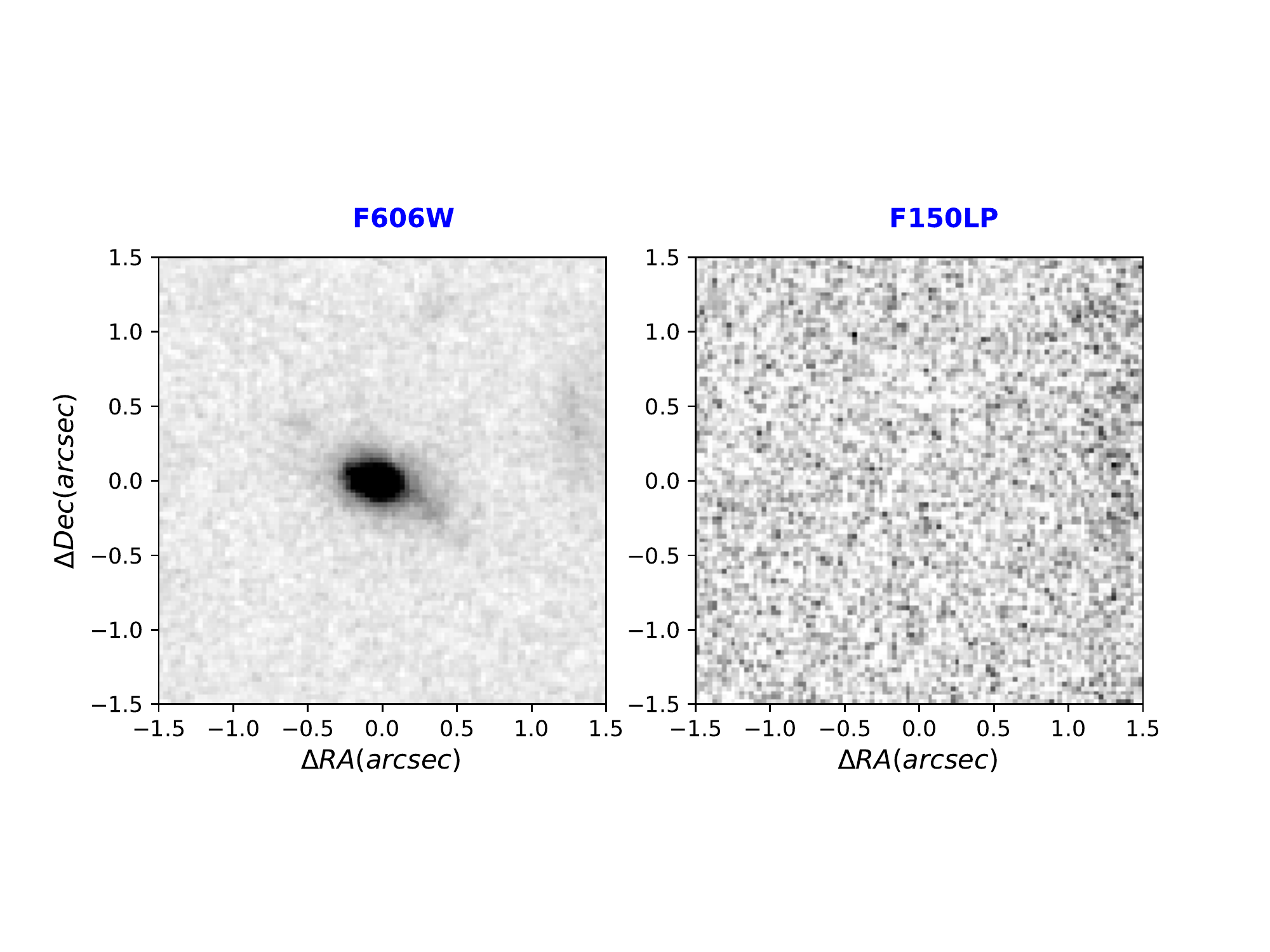}
\caption{ F606W (left) and far-UV F150LP (right) stacked images of our 7 targets. 
We find no detection of LyC flux in the stacked image either.}
\label{fig:stack}
\end{figure}

We stack the far-UV F150LP cutout images to estimate an average escape fraction. To display the stacked image (see Figure \ref{fig:stack}), we 
do a simple addition of F150LP $3\arcsec \times 3\arcsec$ cutouts, centered at the 
position of each source in the F606W image. As shown in Figure \ref{fig:stack}, we do not detect a signal in the stacked image.
We note that because galaxies have various sizes and morphologies, in the case of a simple addition, some galaxies will add noise to areas where other galaxies have flux. 
Therefore, to measure the stacked flux, we perform an optimized stacking \citep[see][]{sia10} by only summing
the pixels that were in the isophotal segmentation of individual galaxy photometry (see Section \ref{sec:photometry}). We also measure the 
total noise in the stack by adding the noise (i.e., $\sqrt{\text{Dark}*\text{Exposure Time}}$) in quadrature in the pixels of individual galaxies that were used in the optimal stacking. The total LyC flux in the stack image is $0.3\times 10^{-31} \ \pm 1.1\times 10^{-31} \ \text{erg} \ \text{s}^{-1}\ \text{cm}^{-2}\ \text{Hz}$. The 3$\sigma$ limit of the stacked flux is reported in table \ref{tab:limits}.


\subsection{Escape Fraction of Ionizing Photons}
\label{subsec:escape fraction}
Lyman continuum radiation produced by young and hot stars is absorbed by neutral hydrogen 
inside galaxies and dust in the ISM, preventing that radiation from escaping the galaxy and reaching the IGM.
There are two broadly-used definitions for the LyC escape fraction in 
the literature. 

First, the {\it absolute escape fraction}, $f_{esc,abs}$, is simply 
the fraction of intrinsic LyC photons that escape into the IGM. This definition is convenient to 
use in theoretical and simulation studies where the true number of LyC photons produced 
is known from the star formation rate and initial mass function. 
 However, this quantity is difficult to measure in observational studies because it requires a 
 measure of the intrinsic production rate of ionizing photons (i.e., LyC luminosity). 
The intrinsic LyC luminosity is usually estimated from nebular emission lines such as 
H$\alpha$ or from the best-fit SED to the galaxy photometry. 
Both of these techniques need an understanding of the dust attenuation (i.e., dust extinction model) and thus they suffer from the associated uncertainties. 
For example, \citet{ste18} show that changing the attenuation relation could change the estimated $f_{esc,abs}$ by a factor of more than three.
 
 The second definition, first introduced by \citet{ste01}, is the {\it relative escape fraction}, $f_{esc,rel}$,
 referring to the fraction of LyC photons that escape the galaxy 
 relative to the fraction of escaping non-ionizing photons at 1500 \AA. A benefit of this quantity is that it is independent of uncertainties in the estimation of dust correction. 
  The $f_{esc,rel}$ can be expressed as below:

\begin{equation}
\label{frel1}
f_{\text{esc,rel}}= \frac{f^{\text{out}}_{\text{LyC}}/L^{\text{int}}_{\text{LyC}}} {f^{\text{out}}_{1500}/L^{\text{int}}_{1500}} = \frac{(f_{\text{LyC}}/f_{1500})^{\text{out}}}{(L_{\text{LyC}}/L_{1500})^{\text{int}}}
 \end{equation}
 
 where the $f^{out}_{\text{LyC}}$ is the LyC flux density per unit frequency in the vicinity of galaxy right after 
 escaping the ISM. Also, $f^{out}_{\text{1500}}$ is the flux density per unit frequency measured at 1500 \AA\ after passing through the galaxy ISM. These flux values are related to the 
 observed fluxes as follows:
 
 \begin{equation}
 \label{frel2}
\left( \frac{f_{\text{LyC}}}{f_{1500}} \right) ^{\text{out}}=\left(\frac{f_{\text{LyC}}}{f_{1500}}\right)^{\text{obs}} \times \text{e}^{\tau_{\text{IGM}}(\text{LyC})}
 \end{equation}
where $\tau_{\text{IGM}}(\text{LyC})$ is the optical depth of LyC photons through the IGM along the 
line of sight to that galaxy. We note that we measure the LyC flux at 700 \AA, while it is usually measured at 900 \AA\ in most 
studies \citep[e.g., ][]{ste01,izo16a, izo16b, mar17,ste18, fle19}. 
Considering current observational facilities (i.e, SBC), we only have access to the ionizing flux at 
shorter wavelengths (700 \AA) at $z\sim1$ \citep[see also][]{sia07,sia10}.

As seen in equations \ref{frel1} and \ref{frel2}, to derive $f_{esc,rel}$, we need to 
estimate the IGM absorption and the amplitude of the intrinsic stellar Lyman break, $(L_{\text{1500}}/L_{LyC})^{\text{int}}$. 
Below, we summarize what we use for each of these quantities.\\

{\it IGM:} The IGM absorption ($\text{e}^{-\tau_{\text{IGM}}}$) is computed from a Monte Carlo simulation described in detail in \citet{sia10} and \citet{ala14}. 
In summary, we create 300 different lines-of-sight through the IGM at different redshifts by 
selecting random hydrogen absorbing systems (i.e., Ly$\alpha$ forest, Lyman limit and damped Ly$\alpha$ systems) 
from the density distribution associated with that redshift. We chose to run this simulation for 300 random lines-of-sight to accurately sample 
the column density and number density distributions of the intervening absorbers.

For this simulation, we adopt the number density and column density distributions of the 
intervening absorbers from the literature \citep{jan06,rao06,rib11,ome13} as explained in detail in \citet{ala14}.
We then take the mean IGM absorption from 300 LOSs for each redshift. The IGM absorption value that we 
used for each galaxy is listed in Table \ref{tab:limits}. 


{\it $(L_{1500}/L_{\text{LyC}})^{\text{int}}$}: This intrinsic flux density ratio depends on the age, star formation history, metallicity and IMF. 
Ideally, we would fit each individual SED with a stellar population model and derive the intrinsic flux decrement across the Lyman break. 
However as shown in \citet{sia10}, the precise SFH is ambiguous and the best-fit SEDs from different SFHs give very different predictions of the intrinsic LyC flux. 
This is mainly because the portion of SED to which we are fitting the photometry is dominated by stars with ages $> 100$ Myr, 
whereas the LyC flux comes from massive stars with ages $< 10 $ Myr. 
Relatedly, \citet{rut16} argue that for a given star formation history, the largest uncertainty in this intrinsic flux ratio is 
introduced by the ignorance of the stellar age and the IMF. They further show that this flux ratio can even be affected by stellar rotation and the choice of stellar template libraries.
In addition, as discussed in \citet{ste18}, depending on the assumed age, metallicity, SFH, IMF and the 
effect of binary evolution of massive stars, stellar population models predict a range of $0.15<(L_{900}/L_{1500})^{\text{int}}<0.75$. 

\begin{figure}
\epsscale{2.0}
\includegraphics[trim=1cm 0cm 2cm 0cm, width=\columnwidth]{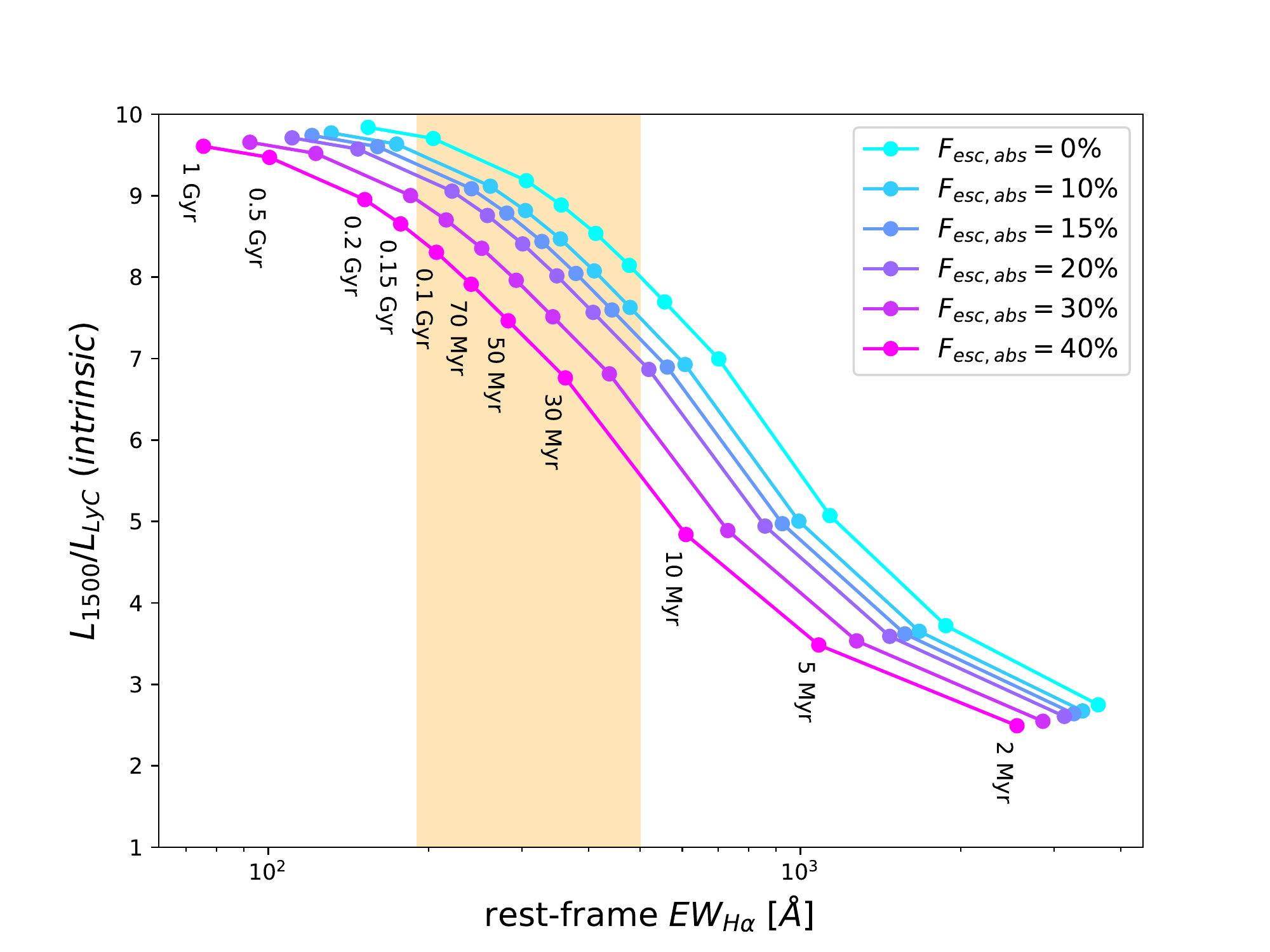}
\caption{The $(L_{1500}/L_{\text{LyC}})^{\text{int}}$ intrinsic flux ratio as a function of rest-frame H$\alpha$ EW using the BC03 models with $Z=0.2 \ Z_{\odot}$ and constant SFH.
 Each color specifies a different $f_{\text{esc,abs}}$ value as listed in the figure. 
 The H$\alpha$ EW of galaxies in the observed sample would be located in the shaded orange region.}
\label{fig:int_ratio}
\end{figure}

Here, we perform a simple analysis to quantify this ratio. We use BC03 \citep{bru03} 
stellar population synthesis models with Chabrier IMF, constant SFH and metallicity of $Z=0.2 \ Z_{\odot}$. We then derive the UV luminosity at two 
wavelengths: the ionizing continuum at 700 \AA\ (i.e., the effective wavelength of the F150LP filter used in our SBC imaging) 
and non-ionizing UV at 1500 \AA\ for a range of ages at 
[2 Myr, 5 Myr, 10 Myr, 30 Myr, 50 Myr, 100 Myr, 150 Myr, 200 Myr, 500 Myr, 1 Gyr], as shown in 
Figure \ref{fig:int_ratio}. At each age, following \citet{ino11}, we derive the H$\alpha$ fluxes and thus the 
H$\alpha$ EWs as below:

\begin{equation}
 \label{equ:halpha}
 \begin{split}
L_{H\beta} &= 4.78 \times 10^{-13}  \frac{1-f_{\text{esc,abs}}}{1+0.6f_{\text{esc,abs}}}  N_{\text{LyC}}\\
L_{H\alpha} &= 2.78 \times L{H_{\beta}} 
\end{split}
\end{equation}
which results from the assumption of Case B recombination \citep{oster06}. The factors in these equations are calculated 
by assuming a temperature $T=10^{4}$ K and an electron density $n_{e}=10^{2}$ cm$^{-3}$.
$N_{\text{LyC}}$ represents the stellar production rate 
of ionizing photons in units of s$^{-1}$. The H$\alpha$ EW values are displayed on the x-axis in Figure \ref{fig:int_ratio}. 
 The galaxies in the sample have H$\alpha$ EWs between 190-500 \AA\ as shown in the 
orange area in Figure \ref{fig:int_ratio}. We note that we are assuming a constant 
SFH here, while an instantaneous SFH would result in the same EW range at younger ages \citep{lei99,amorin15}. From this figure, we see that a $(L_{1500}/L_{\text{LyC}})^{\text{int}}$
ratio between 6 and 10 is required to reproduce the observed EW distribution with reasonable assumptions about the escape fraction. In what follows we use the average value of 8, which is consistent with the values used in some of the previous studies including \citet{sia07} and \citet{rut16}.

 
 Following equations \ref{frel1} and  \ref{frel2} and assuming an intrinsic ratio of 8, 
 we estimate the $f_{\text{esc,rel}}$ values as listed in Table \ref{tab:limits}. In addition, we also calculate the $f_{\text{esc,abs}}$ values because 
 several studies in the literature report their findings in terms of this quantity. To estimate this quantity, we use the relation of 
 $f_{\text{esc,abs}} = 10^{-0.4A_{1500}} f_{\text{esc,rel}} $, as described in detail in \citet{sia07}. 
 Assuming a SMC curve, we derive the dust attenuation at $1500$ \AA, $A_{1500}$, using the A$_{\text{v}}$ estimates (see Table \ref{tab:objects}) from our SED fitting. The $f_{\text{esc,abs}}$ values are listed in Table \ref{tab:limits}. 

As described above, our determination of the escape fraction uses 
the H$\alpha$ EW to estimate the $(L_{1500}/L_{\text{LyC}})^{\text{int}}$ 
ratio with an assumed SFH. However, the assumptions made in this analysis could affect our estimate of the 
$(L_{1500}/L_{\text{LyC}})^{\text{int}}$ ratio, and thus the LyC escape fraction values. 
For example, \citet{sia07} shows that assuming an instantaneous SFH will
 increase the $(L_{1500}/L_{\text{LyC}})^{\text{int}}$ ratio at a given age. We also note that accounting 
 for the binary evolution of massive stars 
will decrease this ratio, as it enhances the late-time ionizing photon production \citep{ste18,fle19}. 
Another important uncertainty is the effect of older stellar populations in our interpretation of H$\alpha$ EW. 
If there is an extremely young and strong burst of star formation on top of an older stellar population from previous bursts, 
then the H$\alpha$ EW will not be as high as expected.
This is because the older stellar populations dominate the rest-optical continuum and thus lower the H$\alpha$ EW. 
To avoid this uncertaity, we can take the $f_{1500}$ out of the equation and 
calculate the absolute LyC escape fraction directly from H$\alpha$ luminosity. If we assume that LyC escape fraction is small, which is a reasonable assumption for our sample, we can then 
derive the total intrinsic LyC luminosity from the H$\alpha$ luminosity assuming that all ionizing photons are absorbed and converted to H$\alpha$ emission line via recombination. We 
use the above BC03 models and compute the conversion from H$\alpha$ luminosity to LyC luminosity to be:

\begin{equation}
 \label{equ:conversion}
L_{700} \ [\text{erg} \ \text{s}^{-1} \ \text{Hz}^{-1}]= 1.7 \times 10^{14}  L_{H\alpha} \ [\text{erg} \ \text{s}^{-1}]
\end{equation}

We calculate the H$\alpha$ luminosity, L$_{H\alpha}$, of our sample by correcting the observed H$\alpha$ fluxes for 
dust attenuation assuming an SMC curve. We then use the above equation to derive the intrinsic LyC luminosity, L$_{700}$, 
for each galaxy in our sample. Finally, we calculate the $f_{esc,abs}$ values using the ratio of the observed LyC flux limits corrected 
for the IGM absorption and the intrinsic $L_{700}$ values from equation \ref{equ:conversion}. We report these new estimates of $f_{esc,abs}$ in Table \ref{tab:limits}. 
These new values are in general agreement with our original $f_{esc,abs}$ estimates.

As explained before in Section \ref{sec:observation}, we observed each galaxy in our sample with a different depth, which 
was calculated using the online HST Exposure Time Calculator to reach a $3\sigma$ limit for $f_{\text{esc,abs}}$ of $5\%$. Our measured $f_{\text{esc,abs}}$
values are consistent with our predicted limits.


\begin{deluxetable*}{lcccccccc}{!t}
\tablecaption{Summary of LyC measurements}
\tablewidth{0pt}
\tablehead{\colhead{ID} & \colhead{U [AB]} & \colhead{$f_{1500}[\mu Jy]$} & \colhead{$f_{\text{LyC}}\tablenotemark{a} [\mu Jy]$} &  
\colhead{IGM transmission (exp(-$\tau$))} & \colhead{$f_{\text{LyC}}/f_{1500}$\tablenotemark{b}}& \colhead{$f_{esc,rel}$\tablenotemark{c}} 
& \colhead{$f_{esc,abs}$\tablenotemark{d}}& \colhead{$f_{esc,abs}$}\tablenotemark{e}} \\
\startdata 
cosmos-1-111    &    23.68     &	  1.23     &	  $<0.015$	&  0.56	&$<0.023$	  &	$<0.18$   &  $< 0.08$	&  $<0.11$	       \\
cosmos-3-69 	&	23.73    &	  1.16     &	  $<0.011$	&  0.56	&$<0.019$	  &	$< 0.15$  &  $< 0.04$ 	&  $<0.06$	       \\
cosmos-3-113   &	24.11    &	   0.83    &  $<0.012$       &  0.56       &$<0.027$	  &	$< 0.22$  &  $< 0.06$	&  $<0.13$	       \\
cosmos-13-80   &	23.95    &	   0.96    &  $<0.008$       &  0.61      &$<0.015$	   &	  $<  0.12$ &  $<0.03$	&  $<0.04$	        \\
cosmos-28-132   &	24.20    &     0.76   &  $<0.008$       &  0.56      &$<0.020$	   &	 $<  0.16$ &  $< 0.07$	&  $<0.06$	       \\
goodss-6-124   &	24.10    &	    0.83   &  $<0.007$       & 0.61      &$<0.014$	    &	  $< 0.11$  &  $< 0.07$	&  $<0.08$	       \\
goodss-27-124   &	24.16    &	    0.78   &  $<0.006$      & 0.61        & $<0.012$         &	$< 0.10$  &  $< 0.06$	&  $<0.04$	       \\
\cline{1-9} \\
stacks	             &    ...     &	  6.54     &	  $<0.033$.    &  0.60	       &  $<0.009 $       &	$<0.07$   &  $< 0.06$	  & ...	     
\enddata 
\tablenotetext{a}{The LyC fluxes are $3\sigma$ limits. }
\tablenotetext{b}{This ratio is corrected for the IGM absorption.}
\tablenotetext{c}{An intrinsic ratio of L$_{1500}$/L$_{\text{LyC}}=8$ is assumed. }
\tablenotetext{d}{These values are derived from adding dust correction to the $f_{esc,rel}$ values.}
\tablenotetext{e}{These values are derived directly from an estimate of intrinsic LyC luminosity using the H$\alpha$ luminosity.}
\label{tab:limits}
\end{deluxetable*}

\section{Discussion}
\label{sec:discussion}

\subsection{Comparison with Other Studies: Physical Properties (SED fitting)}
\label{subsec:sed}
Given the low success rate of identifying LyC leakers at various redshifts, it is important to compare our sample with 
other samples and investigate the LyC escape fraction in the context of different physical properties of galaxies. 

The photometry of our sample results from a combination of ground- and space-based imaging. 
This includes 44 and 40 broad photometric bands from near-UV ($\lambda _{rest} \sim0.19 \ \mu \text{m}$) to IRAC4 
($\lambda _{rest} \sim3.5 \ \mu \text{m}$) in COSMOS and GOODS-South fields, respectively. 
The 3D-HST catalogs provide the best SED fitting parameters for these galaxies, but they assume a metallicity of 1.0 $Z_{\odot}$. However, 
the low stellar masses (i.e., $\log(M^{*}/M_{\odot}) \sim 9-10$) of our sample suggest lower values of [0.2, 0.4] $Z_{\odot}$ for the metallicity 
\citep{wuy12}.
We perform our own SED fitting using the FAST code \citep{kri09} on BC03 
stellar population models assuming a Chabrier IMF \citep{cha03}. We assume an exponentially-increasing star formation 
history (i.e., SFH $\propto e^{t/\tau}$) with $7 < \log(\tau) < 10$, as argued in \citet{red12} for high-redshift galaxies. We allow the metallicity to 
change between two values [0.2, 0.4] $Z_{\odot}$. We also select a SMC extinction curve \citep{gor03} with a range of $0<A_{V}<4$. Some recent studies \citep{,bou16,red18} demonstrate that 
$z=1.5-2.5$ galaxies have an $IRX-\beta$ relation that is consistent with a steep extinction curve similar to SMC. In addition, we correct the 
broadband photometry for the contamination from nebular emission lines using the fluxes measured from the 3D-HST spectra. We note that the 
stellar mass estimates from our SED fitting are close to the 3D-HST stellar masses that we used for our initial sample selection. However, other 
parameters such as dust attenuation snd SFR have significantly changed.


 \begin{figure}
\epsscale{2.0}
\includegraphics[trim=2cm 0cm 2cm 0cm, width=\columnwidth]{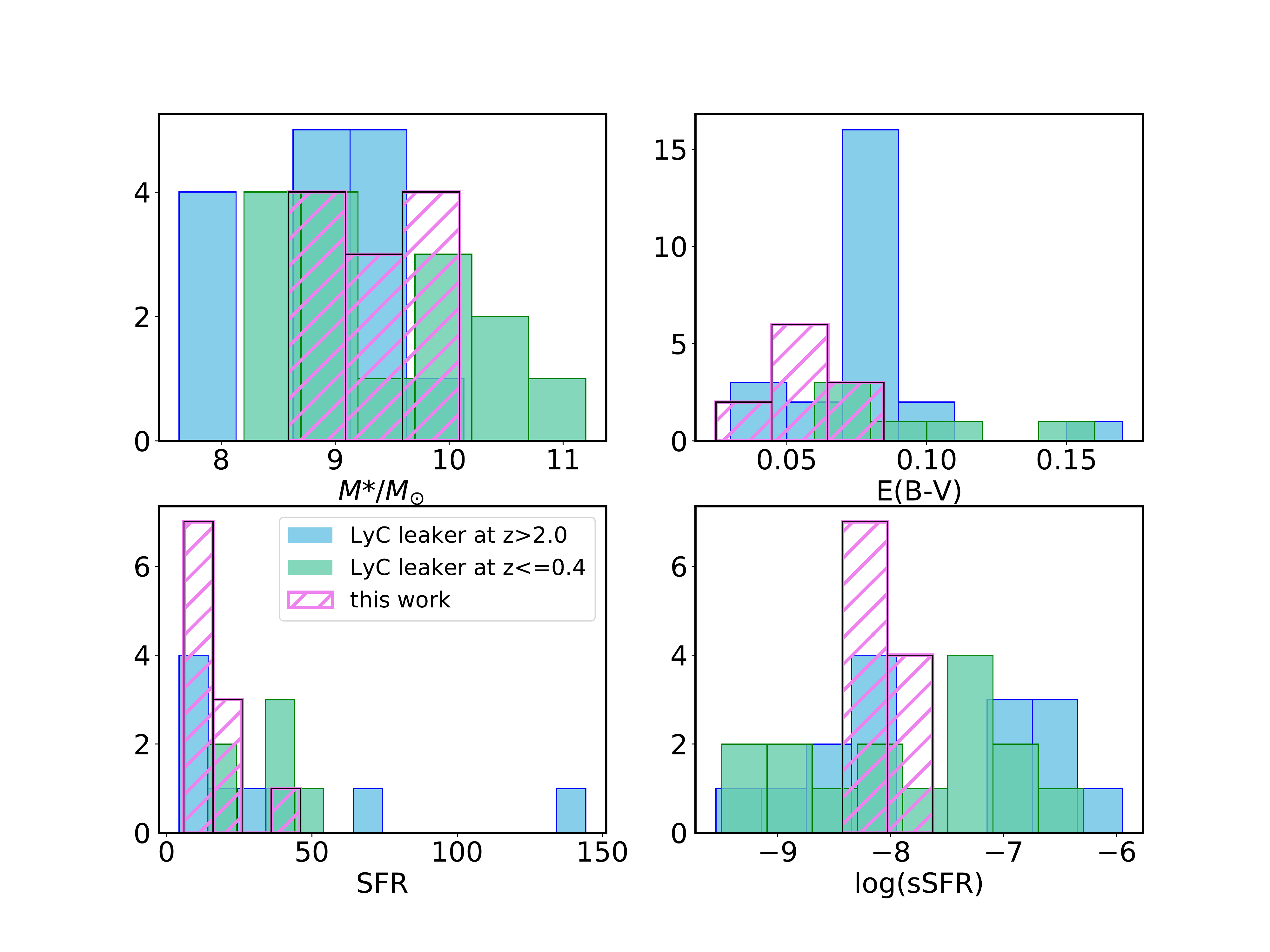}
\caption{ Histograms of the SED fit parameters of M$_{*}$, color excess E(B-V), SFR and sSFR. The confirmed 
LyC leakers from literature are shown in blue and green colors for galaxies 
at $z>2.0$ and $z<0.4$, respectively. The parameter distributions of the galaxies in our sample are shown in hashed pink histograms. 
These galaxies span a similar range of galaxy physical parameters as the other LyC emitters in the literature. }
\label{fig:sed_param}
\end{figure}

 \begin{figure*}[t]
\epsscale{2.0}
\includegraphics[trim=0cm 0cm 0cm 0cm, width=2\columnwidth]{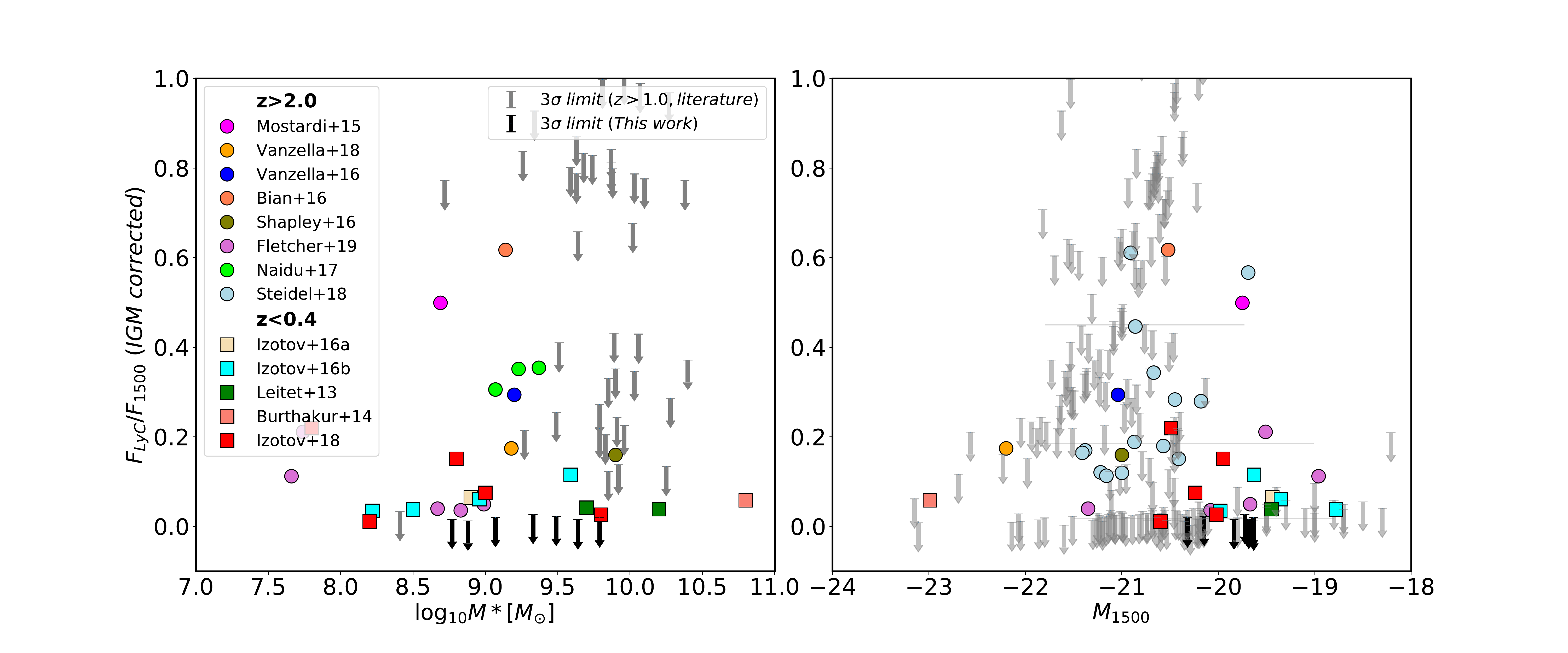}
\centering
\caption{ Left: The observed ratio of $f_{\text{LyC}}/f_{1500}$ corrected for IGM absorption as a function of stellar mass. 
The LyC emitters from the literature are shown with colored circles and squares for 
galaxies at $z>2$ and $z<0.4$, respectively, as listed in the figure. Our $3\sigma$ non-detection limits are shown as black arrows. 
Right: The y-axis is similar to the left panel and the x-axis displays the UV absolute magnitude measured at 1500 \AA. 
For both panels, the $3\sigma$ non-detections from the literature are from 
\citet{sia07,cow09,bri10,sia10, nes13,amo14,sia15,gra16,gua16,rut16,mar17,her18,nai18,smi18}. While the galaxies in our sample 
occupy the same range of M$_{*}$ and $M_{1500}$, their flux ratio limits are lower than the ratio observed for LyC leakers. }
\label{fig:mass_uv}
\end{figure*}

We compare the stellar masses (M$_{*}$), dust reddening parametrized as E(B-V), SFR and specific 
star formation rate (sSFR = SFR$/M^{*}$) of the galaxies in our sample with the quantities of confirmed LyC emitters from
 other studies. We divide the known LyC emitters in two groups of low $z<0.4$ and high redshift $z>2.0$ galaxies. 
 We note that we scale the SFR and sSFR measurements to the Salpeter IMF \citep{sal55} for a
  fair comparison (i.e., multiplying our SFR estimates by a factor of 1.75). 
Caution must be taken when comparing SFR and 
color excess with other studies, because these parameters strongly depend on the dust attenuation models. 
For example, assuming a flat dust curve like \citet{cal00} would result in a larger value of E(B-V). 
Therefore, for the comparisons of E(B-V) and SFR values, we only include the studies that use a similar extinction curve (i.e., SMC). 
In addition, to be consistent with our SFR estimate from the SED fitting, we only compare with the high-redshift studies 
that use an indicator sensitive to recent SFR (i.e., $\sim 100$ Myr) in galaxies \citep{mos15,sha16,van16,bia17,nai17,van18,fle19}. 
However at low redshifts, we do not find recent SFR measurements corrected with the SMC curve in the literature. To compare with studies at $z<0.4$, we use their 
SFR values obtained from extinction-corrected H$\beta$ flux densities reported in \citet{izo16a,izo16b,izo18a,izo18b}. We note that SFRs from nebular emission lines trace
the star formation activity on timescales of 10 Myr.

Figure \ref{fig:sed_param}  compares the distributions of 4 SED parameters of our non-detections (pink hashed histogram) and the corresponding distributions of the other 
LyC leakers at low (green histogram) and high (blue histogram) redshifts. As seen in the upper-left panel of this figure, the stellar mass 
distribution of our non-detections overlaps with the stellar mass range covered by the LyC leakers at low and high redshifts. 
The mean value of our stellar mass distribution $<M^{*}/M_{\odot}>=9.5$ is halfway between the corresponding mean values of $<M^{*}/M_{\odot}>=10$ and 
$<M^{*}/M_{\odot}>=9.1$ for the LyC leakers at low and high redshifts, respectively.


We also examine the distribution of dust attenuation, E(B-V), in the upper-right panel of Figure \ref{fig:sed_param}. 
The dust distribution of our non-detections overlaps with the lower end of the dust distributions for the LyC emitters.

Finally, as shown in the lower panels of Figure \ref{fig:sed_param}, both SFR and sSFR histograms are consistent with the 
corresponding histograms of the LyC emitters. We should reemphasize that the SFR of low redshift LyC galaxies is derived from the H$\beta$ emission line and thus it estimates the 
instantaneous SFR, while SFR estimates of our non-detections and high-redshift LyC galaxies are averaged over a timescale of $\sim 100$ Myr.

In summary, although the confirmed LyC emitters cover a wide range of values for the physical properties discussed above, 
our measured values do fall within these observed distributions. Our observations demonstrate clearly that none of these parameters, alone, can guarantee a LyC-emitting galaxy.


In Figure \ref{fig:mass_uv}, we show the observed ratio of $f_{\text{LyC}}/f_{1500}$ corrected for the IGM absorption 
versus stellar mass (left) and UV absolute magnitude (right) measured at 1500 \AA, $M_{1500}$. The quantity shown on the y-axis 
is the same as $(f_{\text{LyC}}/f_{1500})^{\text{\text{out}}}$ used in equation \ref{frel2}.
 To ensure a fair comparison between these studies and to bring them to the same framework of IGM transmission estimates, 
 we take the value of the observed ratio (or the $3\sigma$ limit) 
 from each paper and correct for IGM absorption by adopting the correction factors at the relevant wavelength from our IGM simulations.
 
  In these plots, we collected measurements of the LyC flux at different redshifts. 
  The confirmed LyC detections are represented in two groups of high-redshift sources at $z>2$ shown with colored circles
 \citep{mos15,sha16,van16,bia17,nai17,ste18,van18,fle19} and low-redshifts sources at $z<0.4$ shown with colored squares 
 \citep{lei13,bor14,izo16a,izo16b,izo18a}. We also include the 
 non-detection $3\sigma$ limits from studies of LyC candidates at $z>1.0$ \citep{sia07,cow09,bri10,sia10,nes13,amo14,sia15,gua16,gra16,rut16,mar17,her18,nai18,smi18}. 
 Our $3\sigma$ limits (black downward arrows) are lower than the ratio measured for the confirmed LyC 
 leakers at higher and lower redshifts. Considering that the galaxies investigated in this study 
 have similar physical properties to the LyC leakers and thus likely the same intrinsic LyC production, our 
 low ratio of $(f_{\text{LyC}}/f_{1500})^{\text{\text{out}}}$ is likely related to the conditions in the ISM, such as 
 high HI column density, and geometrical distribution of dust and neutral gas, which makes it difficult for photons to escape.

 \begin{figure*}[t]
\epsscale{2.0}
\includegraphics[trim=0cm 0cm 0cm 0cm, width=2\columnwidth]{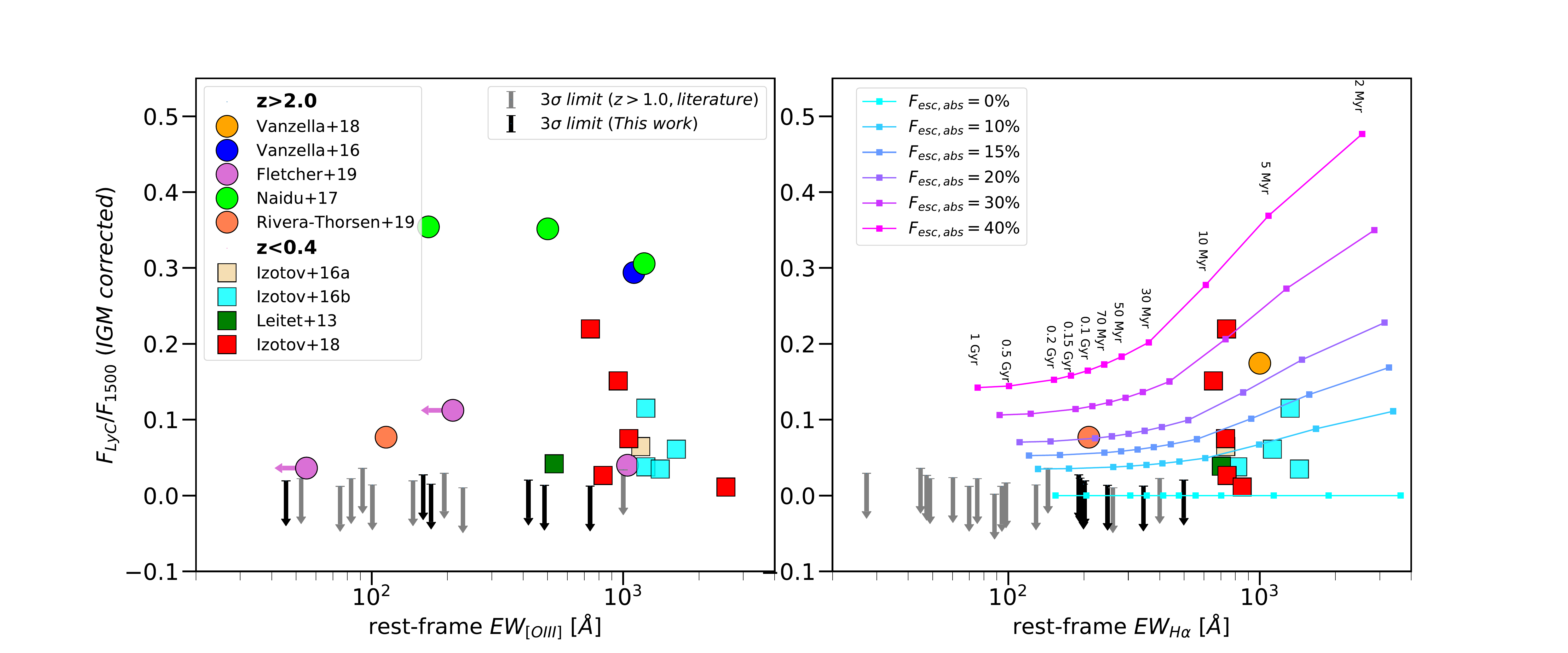}
\centering
\caption{ Left: The y-axis is similar to Figure \ref{fig:mass_uv} and the x-axis shows the rest-frame [OIII] EW. Right: Similar to the left panel 
with the x-axis values representing the H$\alpha$ EW. For both panels, the $3\sigma$ non-detections from the literature are from \citet{amo14,rut16,her18,nai18}. 
The targets in the current work have a distribution of [OIII] EW that is similar to the distribution observed in LyC leakers. 
This conclusion is less clear for the H$\alpha$ EW, where 
the detections have nominally higher H$\alpha$ EW values. One possibility is that the sample of LyC emitters with H$\alpha$ EW 
measurement is somewhat incomplete 
due to lack of observations of the H$\alpha$ line at $z>2.5$, where most of the high-redshift LyC emitters are. 
The solid lines in the right panel are obtained from our 
simple BC03 models, which predict the value of $(f_{\text{LyC}}/f_{1500})^{\text{\text{out}}}$ at ages varying between 
2 Myr and 1 Gyr for a range of LyC escape 
fraction assumptions.}
\label{fig:ha-oiii}
\end{figure*}

\subsection{Comparison with Other Studies: H$\alpha$ and [OIII] EWs}
\label{EW}
Galaxies with intense rest-frame optical emission lines, EW ($\text{H}\alpha \ \text{and/or} \ [\text{OIII}] ) >100$ \AA, are known as 
``Extreme Emission Line Galaxies" \citep[EELG; see][]{ate11}.
At the wavelengths of these nebular emission lines, the EW is indicative of the ratio of current SFR -thus numerous hot O-type stars producing 
ionizing LyC photons- to the integrated past SFR. Therefore, EELGs 
are undergoing a starburst episode with a significant population of new stars. Relatedly, \citet{amorin15} demonstrate that these galaxies are 
dominated by young ($<10$ Myr) star-forming regions. 
In a recent work, \citet{red18}, using an extensive spectroscopic 
survey of star-forming galaxies at $z=1.4-3.8$, show that high-EW galaxies, especially those with high [OIII] EW, 
have both high ionization parameter and ionizing photon production rate (i.e, $\xi_{ion}$). A more recent study by \citet{tan19} shows similar results.
These characteristics have made EELGs ideal objects in which to search for escaping ionizing photons.
	
In addition, some studies \citep{hen15,yan17} discovered that $\sim 70\%-100\%$ of Green Peas, EELGs with high [OIII] EW at low-redshift, 
are strong Ly$\alpha$ emitters. Using Ly$\alpha$ radiative transfer simulations, \citet{ver15} have argued that the detection of Ly$\alpha$ in 
emission from galaxies can be used to identify LyC emitters. Specifically, Ly$\alpha$ profiles can be indicative of LyC-leaking star clusters. They show a Ly$\alpha$ 
spectrum with either an asymmetric redshifted profile with small shift or non-zero L$\alpha$ flux blue-ward of the systematic redshift can be an indicator of escaping LyC photons. 
In addition, some observational studies such as \citet{ste18} and \citet{fle19} found high LyC escape fraction for Ly$\alpha$ emitters. This 
is another piece of evidence that EELGs are likely to be LyC emitters.


 \begin{figure*}[t]
\includegraphics[trim=0cm 0cm 0cm 0cm, width=2\columnwidth]{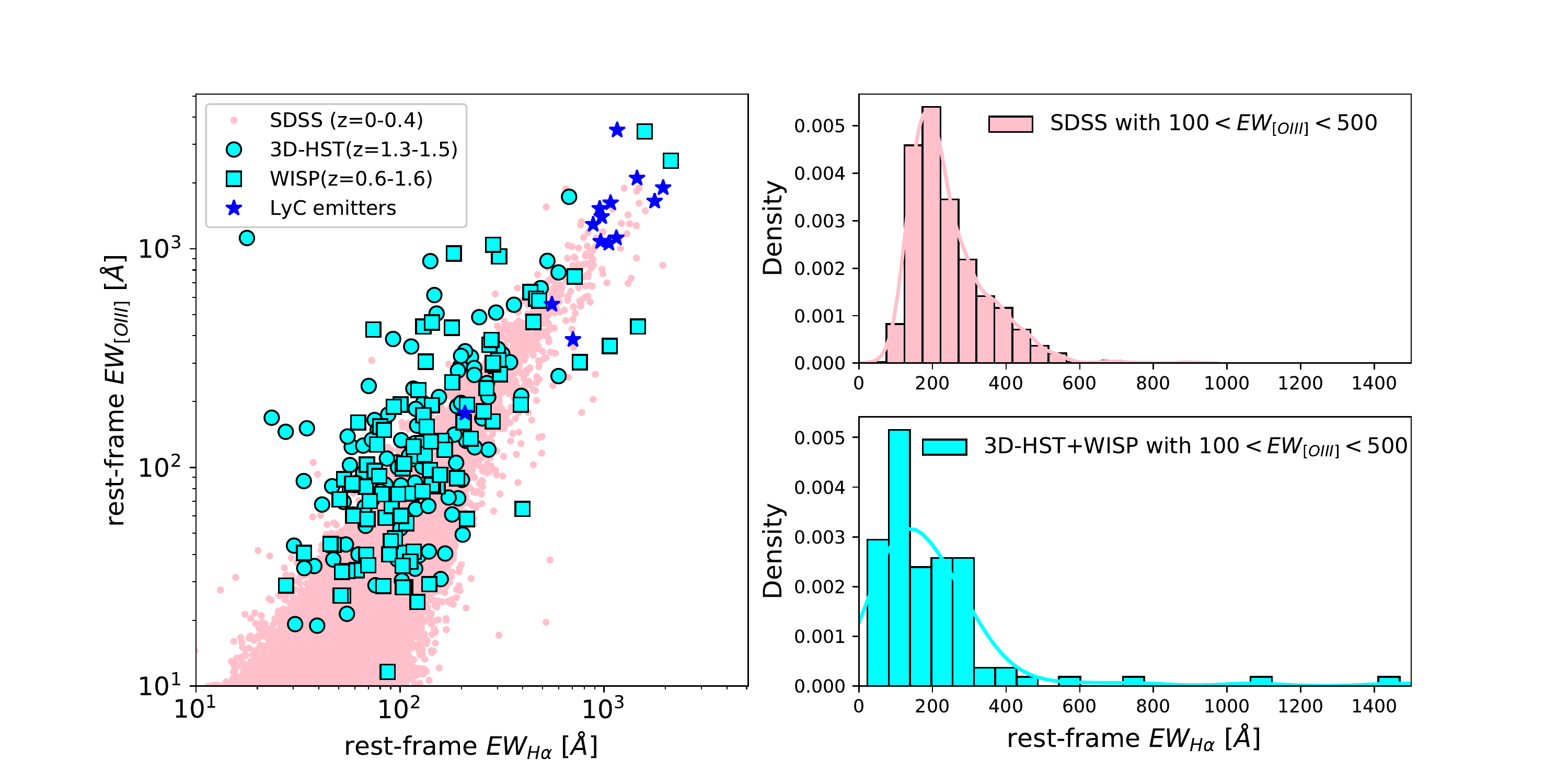}
\centering
\caption{Left: This plot compares the rest-frame H$\alpha$ and [OIII] EW values of galaxies at $1.3<z<1.5$ in the 3D-HST survey. The correlation seen between 
these two quantities allows us to predict the H$\alpha$ EW of the medium-intensity [OIII] emitters. Right (top): The distribution of H$\alpha$ EW of SDSS galaxies with moderate 
[OIII] EW of $100<EW<500$. Right (bottom): The distribution of H$\alpha$ EW of 3D-HST+WISP galaxies with moderate 
[OIII] EW of $100<EW<500$. Both histograms peak around medium values of H$\alpha$ EWs and they do not cover extreme vales of $\sim >600$\AA.   }
\label{fig:ha-oiii_3dhst}
\end{figure*}

The incidence of galaxies with high EW of emission lines (either H$\alpha$ or [OIII] with rest-frame $\text{EW}>100$ \AA) 
in the currently available sample of LyC leakers at low redshifts 
 \citep{izo16a,izo16b,izo18a,izo18b} and high redshifts \citep{van16,nai17,van18,fle19}, suggests that high EW nebular emission lines may be 
a potential indirect tracer of a high escape fraction of ionizing photons. We investigate this 
by comparing the distribution of H$\alpha$ (and [OIII]) EWs of 
the confirmed LyC emitters and the non-detections including EELGs in this study. 

Our sources are selected to have strong H$\alpha$ emission lines and they cover the 
EW range between EW$_{\text{H}\alpha} \ = \ 190-500$ \AA. They also 
have intense [OIII] emission lines with rest-frame EW$_{[\text{OIII}]} \ = \  50-740$ \AA. 
As explained in detail in section \ref{sec:results}, 
we do not detect significant ionizing flux in either the individual galaxies or the stack. 
This is not the first time that observations of EELGs have not detected escaping LyC photons \citep[e.g.,][]{rut16,nai18}. 
Figure \ref{fig:ha-oiii} compiles the LyC measurements, detections and non-detections, 
where an estimate of H$\alpha$ and/or [OIII] EW is available. 
It shows the IGM corrected $f_{\text{LyC}}/f_{1500}$ as a function of [OIII] EW in the left panel and H$\alpha$ EW in the right panel.

{\it[OIII] EW}: As seen in the left panel in Figure \ref{fig:ha-oiii}, galaxies in our sample (black arrows) span a wide range of rest-frame [OIII] EW similar to 
the confirmed LyC leakers at lower and higher redshifts \citep{lei13,nai17,fle19,riv19}. 
However, our sample does not cover the very 
extreme EW values (i.e., [OIII] EW $>1000$ \AA\ ) seen in some of 
the LyC emitters at low \citep{izo16b,izo18a,izo18b} and high redshift \citep{van16}. 
This plot re-confirms the conclusion presented in previous studies \citep{izo17,nai17,fle19} 
that high [OIII] EW on its own 
is an insufficient diagnostic tool for the leakage of LyC photons. 

{\it H$\alpha$ EW}: In the right panel of Figure \ref{fig:ha-oiii}, we investigate the possibility of a correlation 
between detecting LyC photons and the EW of the H$\alpha$ nebular emission line. 
Unfortunately, there are not many high-redshift LyC emitters for which a measurement of H$\alpha$ EW is available. One galaxy at $z=4$ 
from \citet{van18} \footnote{ It should be noted that the EW measurement for this redshift is not from spectroscopic data. Using its broad-band photometry and 
a clear excess in the 3.6$\mu$m flux, they estimate a rest-frame H$\alpha$ EW of 1000 \AA.} 
and one galaxy at $z=2.4$ from \citet{riv19} are the only high-redshift LyC emitters with H$\alpha$ EW measurements.
Therefore, any conclusion based on their small number should be used with caution.
We note that among all of the high-redshift LyC leakers 
in the literature, more than $90\%$ are at $z>3$ where an estimate of the H$\alpha$ EW is very challenging. 

Our sample with moderate H$\alpha$ EW at EW$_{\text{H}\alpha} \ = \ 190-500$ \AA\ spans the same range of rest-frame H$\alpha$ EW as some of the non-detections in the literature \citep{sia10,rut16,her18, nai18}. The H$\alpha$ EW of a confirmed LyC emitter at $z=2.37$ by \citet{riv19} is also within this range. 
In contrast, the group of confirmed LyC emitters (filled squares and orange circle 
in the right panel of Figure \ref{fig:ha-oiii}) with extreme H$\alpha$ EWs  
including several low-redshift sources at $z<0.4$ and only one high-redshift galaxy \citep[$ion3$ at $z=4$ from][]{van18}, 
looks to be separated from the non-detected sources at moderate H$\alpha$ EWs. 
In what follows, we first suggest two scenarios that can explain the lack of LyC emitters seen at moderate H$\alpha$ EWs . 
We then investigate whether H$\alpha$ EW is an effective proxy for LyC emissivity.


\subsubsection{Moderate H$\alpha$ EW} 
\label{subsec:moderate-Ha}
As mentioned above, while searches for emerging LyC photons have yielded null results 
\citep[except the study by][]{riv19} for galaxies with moderate  H$\alpha$ EWs, galaxies with extreme H$\alpha$ EWs usually seem to be LyC emitters. 
This dichotomy (see the right panel of Figure \ref{fig:ha-oiii}) could be due to small number statistics. However, if the dichotomy is real, it could suggest 
that sources with moderate intensity H$\alpha$ emission have a lower LyC escape fraction than that of the sources with extreme H$\alpha$ EW. Below, we discuss 
these two scenarios.

{\it First Scenario:} We show that a moderate H$\alpha$ EW does not necessary imply a null LyC emissivity, and thus the dichotomy 
seen in Figure \ref{fig:ha-oiii} might be due to small number statistics. As pointed out before, the number of confirmed high-redshift LyC emitters 
with H$\alpha$ EW measurements available is only two compared to the eight high-redshift LyC emitters with [OIII] EW 
measurements discussed above. Here, we perform a validation experiment to 
predict the H$\alpha$ EWs of those LyC emitters without H$\alpha$ line measurements and to investigate if any of them are likely to have moderate H$\alpha$ EW. 

 
We select 3D-HST galaxies at $1.3<z<1.5$, where both H$\alpha$ and [OIII] emission lines are available. We 
re-measure the EW values of these two lines, because the 3D-HST catalogs have been reported to overestimate the EWs when the continuum detected by the 
grism is faint \citep{nai17}. In addition, we include galaxies from the WFC3 Infrared Spectroscopic Parallel Survey \citep[WISP,][]{ate10}. 
The WISP survey obtained slitless near-IR grism spectroscopy of more than 200 independent fields in pure-parallel mode with WFC3. This survey
observes both H$\alpha$ and [OIII] emission lines for galaxies at $0.6<z<1.6$. For comparison, we also
include a sample of $\sim100,000$ local galaxies at $z < 0.4$ drawn from the Sloan Digital Sky Survey \citep[SDSS,][]{york00} DR12 release 
\citep{ala15}.

As shown in the left panel of Figure \ref{fig:ha-oiii_3dhst}, the 
H$\alpha$ and [OIII] EWs strongly correlate such that we do not expect extreme H$\alpha$ EW (i.e., > 600 \AA) for moderate-intensity [OIII] lines 
(i.e., $100<\text{EW}_{[\text{OIII}]}<500$). To 
further investigate this quantitatively, we plot the distribution of H$\alpha$ EW values for a range of moderate [OIII] EW values of $100<\text{EW}_{[\text{OIII}]}<500$ for 
SDSS and 3D-HST+WISP galaxies in the upper-right and lower-right panels of Figure \ref{fig:ha-oiii_3dhst}, respectively. As seen in these histograms, 
galaxies with moderate [OIII] EW values are very likely to be moderate H$\alpha$ emitters. Therefore, those
LyC emitters with moderate [OIII] EW values (e.g., purple and light green circles in the left panel of Figure \ref{fig:ha-oiii}) and no H$\alpha$ line measurement, are presumably moderate H$\alpha$ emitters as well.

We should note that the [OIII] emission line is sensitive to the presence of 
very young hot stars. H$\alpha$ is instead sensitive to the presence of somewhat less extreme photons (i.e., 13.6 eV). Therefore, 
for a given moderate-intensity [OIII] line, it is theoretically possible to obtain an extreme H$\alpha$ EW. However, as seen 
in Figure  \ref{fig:ha-oiii_3dhst}, observational values from SDSS, 3D-HST and WISP surveys do not show many of such examples, and thus they must be rare.

In addition, we over-plot the LyC emitters (blue stars) from literature in the 
left panel of Figure \ref{fig:ha-oiii_3dhst}. As illustrated in this figure, the currently known LyC emitters follow the mean relation between the [OIII] and H$\alpha$ EWs 
seen in the SDSS, 3D-HST and WISP galaxies. Therefore, it is likely that the absence of moderate H$\alpha$ emitters is just a selection effect.

Finally, we would like to emphasize that this scenario and its 
conclusion are solely based on the prediction of H$\alpha$ EW values, and a definitive answer requires direct measurement of H$\alpha$ 
emission lines for the LyC emitters with moderate [OIII] EWs.

{\it Second Scenario:} The lack of galaxies with high LyC emissivity at moderate H$\alpha$ EWs may be indicative of a real relation. 
To further investigate this possibility, we use our simple models (as described in Section \ref{subsec:escape fraction}) to understand how the observed ratio of 
$f_{\text{LyC}}/f_{1500}$ changes with H$\alpha$ EW. We calculate the observed $f_{\text{LyC}}/f_{1500}$ ratios in 
our models by applying a range of assumed escape fraction values, 
$f_{esc,abs}=[0\%,10\%,20\%,30\%,40\%]$, to the intrinsic $f_{\text{LyC}}/f_{1500}$ ratios calculated in Section \ref{subsec:escape fraction}. To be consistent with 
the observed ratios of the real galaxies, we also add the effect of dust attenuation at LyC and $1500$ \AA\ wavelengths, using a median of
the color excess values, E(B-V), of all the LyC emitters in the literature. As illustrated in the right panel of Figure \ref{fig:ha-oiii}, the observed ratios are calculated at
different ages, ranging from 2 Myr to 1 Gyr, corresponding to different H$\alpha$ EWs.

The lines seen in the right panel of Figure \ref{fig:ha-oiii} are the results of variation of age and LyC escape fraction 
in the models. The age parameter is changing in the direction of the x-axis such that stellar populations are aging as we move along these lines toward lower H$\alpha$ EWs. 
On the other hand, the LyC escape fraction parameter causes the spread in the direction of the y-axis. 
Based on our simple models, we can think of galaxies with moderate H$\alpha$ EWs to be similar to galaxies with extreme H$\alpha$ EWs but with older ages. Therefore, if we 
assume that the distribution of LyC escape fraction at moderate H$\alpha$ EWs is the same as the distribution seen at extreme H$\alpha$ EWs, then we would expect to 
detect some LyC emitters at lower H$\alpha$ EWs. Because we do not detect any LyC emission at these moderate H$\alpha$ EWs, we can conclude that the 	LyC escape 
fraction in older galaxies (i.e., with moderate H$\alpha$ EWs) must be lower.


In conclusion from the two scenarios described above, whether the lack of LyC emitters at moderate H$\alpha$ EWs is due to small 
number statistics (i.e., first scenario in which we could see both LyC emitters and non-detections) or having a lower LyC escape fraction (i.e., second scenario in which 
we mostly see non-detections), moderate H$\alpha$ EW, alone, is not a promising indicator of lyC leakage.

\subsubsection{Extreme H$\alpha$ EW}
With the currently available data (i.e., finding only LyC emitters and no non-detections at extreme 
H$\alpha$ EWs), we can suggest that an extreme H$\alpha$ emission 
(i.e., H$\alpha$ EW $ \sim 600-1000$ \AA\ and beyond) is likely an effective tracer of LyC emissivity.

However, we would like to note the work by \citet{izo17} who indirectly derive 
the LyC escape fraction of a sample of local compact star-forming galaxies with extreme H$\alpha$ and [OIII] emission lines. They investigate whether 
the galaxies in their sample can emit LyC by constraining their neutral gas column densities through photoionized H II region models.  
For two out of five of the galaxies in their sample with extreme 
H$\alpha$ emission (H$\alpha$ EW $\sim 2000$ \AA , private communication), they derive a high neutral gas column density, implying a negligible LyC escape fraction. 
In case of a direct observation, this negligible LyC escape fraction would likely result in null LyC emissivity. This would 
contradict our suggestion that an extreme H$\alpha$ emitter is likely a LyC leaker. More definite conclusion about these sources will require 
a direct observation of their LyC emission.

 Finally, we should emphasize that the above discussions ignore the effect of redshift evolution on the properties of LyC emitters. 
 As seen in Figure \ref{fig:ha-oiii}, all of the low-redshift LyC emitters have extreme [OIII] and H$\alpha$ EWs (i.e., $>600$ \AA), while the high-redshift LyC emitters cover a wider 
 range of EW values. A deeper understanding of possible redshift evolution requires a larger sample of confirmed LyC emitters at various redshifts.



\subsection{Understanding the Non-detections}
Our SED fits show that the physical properties (stellar mass, SFR and dust attenuation) of the galaxies in our sample are 
similar to those of the confirmed LyC emitters. 
However, the galaxies in our sample have lower H$\alpha$ EWs than almost all of the LyC emitters with 
H$\alpha$ line measurements in the literature 
\citep[with an exception of a study by][]{riv19}. 
A direct interpretation of our results is that the galaxies in our sample are older and thus they are likely to have 
lower LyC escape fraction (see Section \ref{EW} and the right panel of figure \ref{fig:ha-oiii}). This simple interpretation 
explains why we did not detect escaping LyC emission in this study.

The above interpretation simply explains our non-detections in comparison with the LyC 
emitters with measured H$\alpha$ EWs as seen in Figure \ref{fig:ha-oiii}. However, 
we saw (see Section \ref{subsec:moderate-Ha}) that this plot is probably more complicated than it looks. To further understand our non-detections 
in the context of these possible complications, we discuss other scenarios below.

While LyC emitters with H$\alpha$ line measurements have extreme H$\alpha$ EWs, we showed that LyC emitters with moderate H$\alpha$ EWs possibly exist 
as well (see Section \ref{subsec:moderate-Ha}). If we accept such a possibility, we need to understand why the galaxies in our sample, with H$\alpha$ EWs similar to those 
possible LyC emitters, have low LyC emissivity.

Because the stellar populations, 
and thus the LyC photon production of the galaxies in our sample and those possible LyC emitters at moderate H$\alpha$ EWs are the same, the non-detections in our sample must be 
a result of the LyC photons not being able to escape. To further understand this, we consider a scenario suggested by hydrodynamical simulations \citep{cen15,ma15,paar15}, 
in which the ionizing photons escaping from galaxies into the IGM are highly anisotropic. Consequently, as also noted in \citet{paar15}, 
even if the galaxies in our sample have high {\it actual} escape fraction, there can be many sight lines through which
no ionizing radiation escapes. Therefore, our results strongly depend on the orientation of the galaxies. This emphasizes the importance of including large samples in the observational studies.  

We note that the direction dependency of LyC escape fraction seems to be more significant at lower H$\alpha$ EWs than is at extreme H$\alpha$ EWs. 
As seen in Figure \ref{fig:ha-oiii}, 
all LyC searches at extreme H$\alpha$ EWs resulted in LyC detections and there is no non-detections at these extreme EWs, in contrast to what we would expect for an anisotropic distribution of escaping LyC emission. To 
better understand this conflict, we consider a ``picket fence'' model in which the neutral gas surrounding the ionizing sources is patchy. In this model where parts of the galaxy are
covered with optically thick clouds, LyC photons escape from optically thin holes between the clouds. The distributions of these holes and their sizes likely depend on how strong the 
galactic feedback is. In an extreme starburst (i.e., extreme H$\alpha$ EW), strong stellar feedback may effectively expel neutral gas and lower the covering fraction, thus allowing LyC photons to escape more easily in 
different directions. However in a less extreme condition (i.e., moderate/low H$\alpha$ EW), the covering fraction is higher and weaker stellar feedback creates random optically thin channels in the ISM \citep{zac13}. 
In this situation, we can detect the LyC photons only if they are emitted in the thin channels directed toward our line of sight.

Future observations of the LyC escape fraction will require a large, representative sample of galaxies. To achieve this goal, effective selection techniques will be vital.
As is also discussed in 
\citet{rut16} and \citet{fle19}, we are lacking very low-mass galaxies (i.e., $\log (\text{M}_{*}/\text{M}_{\odot}) < 9$), the so-called dwarf galaxies, in the LyC studies at high redshifts including 
this study at $z \sim 1$. As suggested in many theoretical studies and most simulations \citep[e.g., ][]{yaj11,wise14,lew20,ma20}, these low-mass galaxies likely have a high escape fraction of ionizing photons. 
For example, a recent work by \citet{ma20} predicts that LyC escape fraction increases with stellar mass up to $M^{*} \sim 10^{8} M_{\odot}$ and decreases at higher masses. 
Therefore, a likely path for future studies of LyC escape fraction at $z \sim 1$ is to observe very low-mass dwarf galaxies as promising candidates for escaping ionising radiation.

\section{Summary}
We have obtained ACS/SBC far-UV imaging of 11 star-forming galaxies at $1.2<z<1.4$ to search for escaping LyC emission. We select our targets from the 3D-HST survey to 
have strong H$\alpha$ emission lines with EW $> 190$ \AA\ and low stellar masses with $\log (\text{M}_{*}/\text{M}_{\odot}) < 10$. These criteria identify sources that are undergoing a 
starburst episode with a significant population of new stars and thus they are ideal to search for escaping ionizing photons. Our findings are as follows.

\begin{itemize}

\item After careful data reduction and subtraction of the dark current as the dominant 
source of noise, we do not detect (i.e., S/N $ >3$) any escaping LyC radiation from the individual targets or in the stack.

\item We run a Monte Carlo simulation \citep{sia10,ala14} to compute the IGM absorption. This simulation properly accounts for varying opacity of the IGM along different lines of sight. Applying these IGM corrections, we calculate $3\sigma$ limits of $f_{\text{LyC}}/f_{1500}$ $< [0.014-0.027]$ and $< 0.009$ for the individual 
galaxies and stack, respectively. Assuming an intrinsic ratio of $(L_{1500}/L_{\text{LyC}})^{\text{int}}=8$, these limits translate to 
$3\sigma$ limits of $f_{esc,rel} < [0.10,0.22]$ and $f_{esc,rel}<0.07$ for the individual galaxies and stack, respectively.

\item We fit stellar population models to the multi-band photometry of our sample covering from rest-frame UV to near-IR to estimate the physical parameters. 
The galaxies in our sample exhibit similar ranges of stellar mass, SFR and dust attenuation values as the confirmed LyC emitters in the literature. Our findings indicate 
that none of these galaxy parameters, alone, is a promising indicator of LyC leakage.

{\item We compare the H$\alpha$ and [OIII] EW values of our sample and those of the confirmed LyC emitters in the literature. 
Our H$\alpha$ and [OIII] EW estimates are in the range of EW$_{rest} \ \sim\ [190-500]$ \AA\ and $[50-700]$ \AA, respectively. 
Our sample does not cover extreme values (i.e., $>600$ \AA) seen in some of the LyC emitters in the literature \citep{izo16a,izo16b, van16, izo18a, izo18b,van18}. 
For [OIII] emission lines, we find that high [OIII] EW values do not guarantee the detection of LyC flux. This 
conclusion is consistent with the findings from previous studies \citep{izo17,nai17,fle19}. For the H$\alpha$ emission line, we demonstrate that a moderate EW 
(i.e.,  $< 600$ \AA) is not a promising indicator of leaking LyC photons. 
However, considering current evidence, it is likely that extreme H$\alpha$ emission 
(i.e., $\sim 600-1000$ \AA\ and beyond) is an effective indicator of LyC leakage.}

\end{itemize}

Future observations of LyC emission may have to combine various indirect selection techniques (i.e., profile of Ly$\alpha$ line, UV spectral slope, ultra-faint dwarf 
galaxies with very low stellar mass) to better identify LyC emitters at of $z \sim 1$.

\section{Acknowledgement}
The primary data for this work were obtained with the Hubble Space Telescope operated by AURA, Inc., for NASA under contract NAS 5-26555. Support for this work was provided by NASA through grant HST-GO-14123 from the Space Telescope Science Institute, which is operated by AURA, Inc., under NASA contract NAS 5-26555. Some of the data presented in this paper were obtained from the Mikulski Archive for Space Telescopes (MAST). STScI is operated by the Association of Universities for Research in Astronomy, Inc., under NASA contract NAS5-26555. Support for MAST for non-HST data is provided by the NASA Office of Space Science via grant NNX13AC07G and by other grants and contracts. 

\facility{HST (ACS), MAST (HLSP)}

\bibliographystyle{apj}
\bibliography{citation.bib}

\end{document}